\documentclass[aps,twocolumn,preprintnumbers,amsmath,amssymb,superscriptaddress,longbibliography]{revtex4-1}
\usepackage{subfigure}
\usepackage{bbold}
\usepackage{graphicx}
\usepackage{dcolumn}
\usepackage{color}
\usepackage{amsmath}
\usepackage[breaklinks,colorlinks,bookmarks=false,citecolor=red,linkcolor=blue,urlcolor=blue]{hyperref}
\usepackage{braket}
\usepackage[utf8x]{inputenc} 
\begin{document}

\preprint{\href{https://doi.org/10.1103/PhysRevB.102.035160}{Phys. Rev. B {\bfseries 102}, 035160 (2020)}}

\title{Spin-caloritronic transport in hexagonal graphene nanoflakes}

\author{Th\d{i} Thu Ph\`ung}
\affiliation{Laboratoire de Physique Th\'eorique et Mod\'elisation, CNRS UMR 8089,
  CY Cergy Paris Universit\'e, 95302 Cergy-Pontoise Cedex, France}
\affiliation{Department of Advanced Materials Science and 
   Nanotechnology, University of Science and Technology of Hanoi, 18 Hoang 
   Quoc Viet, 100000 Ha Noi, Vietnam}

\author{Robert Peters}
\affiliation{Department of Physics, Kyoto University, Kyoto 606-8502, Japan.} 

\author{Andreas Honecker}
\affiliation{Laboratoire de Physique Th\'eorique et Mod\'elisation, CNRS UMR 8089,
  CY Cergy Paris Universit\'e, 95302 Cergy-Pontoise Cedex, France}
  
\author{Guy Trambly de Laissardi\`ere}
\affiliation{Laboratoire de Physique Th\'eorique et Mod\'elisation, CNRS UMR 8089,
  CY Cergy Paris Universit\'e, 95302 Cergy-Pontoise Cedex, France}

\author{Javad Vahedi}
\affiliation{Laboratoire de Physique Th\'eorique et Mod\'elisation, CNRS UMR 8089,
  CY Cergy Paris Universit\'e, 95302 Cergy-Pontoise Cedex, France}
 \affiliation{Jacobs University, School of Engineering and Science, Campus Ring 1, 28759 Bremen, Germany}
\affiliation{Department of Physics, Sari Branch, Islamic Azad University, Sari 48164-194, Iran}

\date{March 2, 2020; revised May 26, 2020}

\begin{abstract}
We investigate the spin-dependent thermoelectric effect of graphene flakes 
with magnetic edges in the ballistic regime. Employing static, 
respectively, dynamic mean-field theory we first show that magnetism 
appears at the zigzag edges for a window of Coulomb interactions that 
increases significantly with increasing flake size. We then use the 
Landauer formalism in the framework of the non-equilibrium Green's function 
method to calculate the spin and charge currents in magnetic hexagonal 
graphene flakes by varying the temperature of the junction for different 
flake sizes. While in non-magnetic gated graphene the temperature gradient 
drives a charge current, we observe a significant spin current for 
hexagonal graphene flakes with magnetic zigzag edges. Specifically, we 
show that in the ``\textit{meta}'' configuration of a hexagonal flake 
subject to weak Coulomb interactions, a pure spin current can be driven 
just by a temperature gradient in a temperature range that is promising 
for device applications. Bigger flakes are found to yield a bigger window 
of Coulomb interactions where such spin currents are induced by the 
magnetic zigzag edges, and larger values of the current.
\end{abstract}

\maketitle
\section{Introduction}

The thermoelectric effect, \textit{i.e.}, the direct conversion of a 
temperature difference to an electric voltage and vice versa, attracts 
great attention in recent years with the development of electronics and 
spintronics. Many investigations have addressed the fundamental physics 
and potential applications of thermoelectric phenomena,
see, e.g., Refs.~\cite{Kirihara2012,Kim2014}. 
With shrinking the size of electronic devices to molecular-scale 
electronics~\cite{Aviram1974} beyond the foreseen Moore's law limits of 
small-scale conventional silicon integrated circuits, heat dissipation 
becomes a severe problem due to a high energy consumption~\cite{Pop2006}. 
On the one hand, converting the dissipated heat to electric energy via the 
thermoelectric effect is an interesting solution to this problem. On the 
other hand, in the field of spintronics, the coupling of spin and charge 
transport provides another excellent possibility to reduce the energy 
dissipation in nanoscale devices~\cite{Wolf2001,Fert2008}; the emergent 
field of \textit{spin caloritronics} promises new functionality exploiting 
the interplay of spin and heat currents~\cite{Uchida2008,Adachi2013}. Many 
works in this field have shown that spin caloritronics would be a viable 
scheme to realize low power consumption in molecular-scale 
electronics~\cite{Uchida2010,Jaworski2010,Bauer2012}.

In recent years, graphene has attracted a tremendous amount of attention, 
mostly due to its peculiar electronic structure with massless Dirac cones 
at the Fermi level~\cite{CastroNeto2009,Sarma2011,Novoselov2012}. Graphene 
is generally believed to be a semimetal with at most weak electronic 
correlations. Nevertheless, intriguing theoretical investigations made 
even before single-layer graphene was isolated for the first time 
\cite{Novoselov04} predicted that states localized at zigzag edges give 
rise to magnetic instabilities even when the bulk is semimetallic 
\cite{Fujita96,Wakabayashi98,Wakabayashi99}. This edge magnetism arises 
thanks to electronic states that are localized close to a zigzag edge 
\cite{Nakada96,Brey06} and the fact that only one of the two graphene 
sublattices participates in a zigzag edge, thus favoring a ferromagnetic 
alignment of the resulting magnetic moments.

Graphene's distinguished thermal and electronic performance render it one 
of the most outstanding candidates in spin caloritronics. Many 
experimental and theoretical works have investigated the thermally-induced 
spin-transport properties of 
graphene~\cite{Zuev2009,Wei2009,Checkelsky2009,Ghahari2016,Vahedi2016,Bretin2015,Chico2017,Zeng2011,Zhao2012,Ni2013,Chen2014,Liang2012,Farghadan2018,Zberecki2013,Zhai2014,Li2016}. 
Some of these 
works~\cite{Zuev2009,Wei2009,Checkelsky2009,Ghahari2016,Vahedi2016,Bretin2015,Chico2017} 
focused on the electronic properties of graphene. In particular, the sign 
inversion behavior of graphene's thermoelectric power (TEP) across the 
charge neutrality point and the gate dependence of the TEP have been measured 
by Zuev \textit{et al.}~\cite{Zuev2009}. The Seebeck coefficient and the 
Nernst coefficient have been explored in multiprobe graphene junctions, 
and both of them show oscillating behavior versus gate 
voltage~\cite{Wei2009,Checkelsky2009}. The effects of a substrate on the 
TEP have also been explored by depositing graphene on boron nitride to 
suppress the disorder, which has led to a large enhancement of the 
TEP~\cite{Ghahari2016}. The charge and spin Seebeck effects in 
ferromagnetic graphene have been studied by one of the present authors, 
showing that a pure spin current with a large spin figure of merit is 
attainable by varying the spin splitting, temperature, and doping of the 
junction~\cite{Vahedi2016}. Further studies concern rectangular 
rings~\cite{Bretin2015} and bilayer graphene flakes~\cite{Chico2017}.

Another part of the aforementioned 
investigations~\cite{Zeng2011,Zhao2012,Ni2013,Chen2014,Liang2012,Farghadan2018,Zberecki2013,Zhai2014,Li2016} 
focused on the effect of the magnetism arising at zigzag edges 
\cite{Fujita96,Wakabayashi98,Wakabayashi99,Yazyev2010} on transport 
properties. In particular, the spin Seebeck effect in graphene nanoribbons 
has been explored by first-principles calculations~\cite{Zeng2011,Ni2013} 
and mean-field theory~\cite{Zhao2012}. Other geometries that have been 
considered include nanoribbons with sawtooth-modulated 
edges~\cite{Chen2014}, nanowiggles~\cite{Liang2012}, trapezoidal-shaped 
nanoribbons~\cite{Farghadan2018}, nanoporous~\cite{Dong2019}, and also 
doped ferromagnetic zigzag graphene nanoribbons~\cite{Song2020} exhibiting 
interesting effects on the spin thermopower and an enhancement of the 
figure of merit. The spin-dependent Seebeck effect has also been studied 
in other lattices with a honeycomb structure, such as silicene
nanoribbons~\cite{Zberecki2013}, $\alpha$-zigzag 
graphyne nanoribbons~\cite{Zhai2014}, graphene-based magnetic molecular 
junctions~\cite{Li2016}, and magnetic carbon-based organic 
chains~\cite{Tan2020}.

An important class of systems is given by hexagonal zigzag-edges graphene 
nanoflakes (ZGNFs) where the magnetic polarization alternates between 
neighboring 
edges~\cite{Fernandez2007,Bhowmick2008,Viana2009,Feldner2010,*FeldnerE,Roy2014,Valli2016}. 
Recently, Valli \textit{et al.}~\cite{Valli2018,Valli2019} showed that the 
magnetic edges can yield a nearly complete spin polarization of the 
current and proposed ZGNFs as efficient spin-filtering devices.

\begin{figure}[t!]
\centering
\includegraphics[width=0.9\columnwidth]{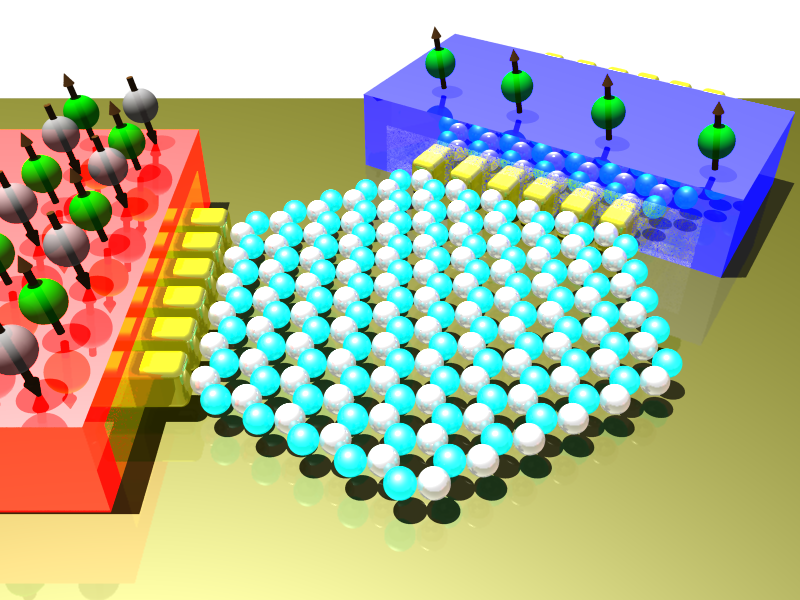}
\caption{Cartoon showing a hexagonal
zigzag-edges graphene nanoflake (ZGNF) attached to semi-infinite metallic leads
to which a temperature difference $\Delta T$ is applied. Left and right lead are
considered to be hot and cold, respectively. Sublattices $A$ and $B$ are drawn in cyan and white color, respectively. In this cartoon, metallic leads are connected to the
sublattice $B$ at different edges in analogy to the \textit{meta} configuration of a benzene molecule.} 
\label{fig1}
\end{figure}

In this paper, we propose a spin-caloritronics device based on hexagonal 
ZGNFs bridged between two non-magnetic metallic leads. In this scheme, 
metallic leads are coupled to the same atomic sublattices of graphene, in 
the same manner as in the \textit{meta} configuration of a benzene 
molecular junction~\cite{Meisner2012,Manrique2015,Wang2019}, see 
Fig.~\ref{fig1} for a sketch. We demonstrate that, by applying a 
temperature gradient between the two leads, pure spin currents 
can be established whose magnitude depends on the reference temperature 
and the gate voltage. We note that such a spin-energy off-diagonal
effect is very different from the aforementioned spin-filtering
effect \cite{Valli2018,Valli2019}. Nevertheless, the 
main ingredient underlying both of these spin-transport effects
are the electronic correlations giving rise to magnetism at 
the zigzag edges in hexagonal 
ZGNFs~\cite{Fernandez2007,Bhowmick2008,Viana2009,Feldner2010,*FeldnerE,Roy2014,Valli2016}.

The remainder of this paper is organized as follows. Section \ref{secII} 
introduces the model and the formalism. In Sec.~\ref{secIII} we present 
and discuss numerical results. Finally, Sec.~\ref{secIV} provides a 
summary and conclusions.

\section{Model and formalism}
\label{secII}

Here we study a hexagonal ZGNF attached to conducting metallic leads. The 
leads are connected to atoms at the edges belonging to the same 
sublattice, as shown with white color in Fig.~\ref{fig1}. Moreover, we 
consider a temperature gradient $\Delta T$ between the two leads.

\subsection{Model}
\label{sec:Model}

The total Hamiltonian describing the device is given by 
\begin{equation}
\mathcal{H} = \mathcal{H}_{L}+\mathcal{V}_{L}+\mathcal{H}_C+\mathcal{H}_{R}+\mathcal{V}_{R},
\end{equation}
where
$\mathcal{H}_{L/R}=\sum_{\alpha k \sigma}c^{\dagger}_{\alpha k\sigma}c_{\alpha
k\sigma}$ corresponds to the left (right) metallic lead. $c^{\dagger}_{i\sigma}$ ($c_{i\sigma}$) creates (annihilates) an electron at site $i$ in the lead $\alpha=L/R$. The term $\mathcal{V}_{L/R}= \sum_{\alpha i k \sigma} \big(V_{\alpha i k \sigma}c^{\dagger}_{\alpha k\sigma}a_{i\sigma}+h.c\big)$ describes the coupling between the central region and the leads, where $V_{\alpha i k \sigma}$ denotes the hopping amplitude between site $i$ of the central region and state $k$ of lead $\alpha=L/R$. The term $\mathcal{H}_C$
describes the central region for which we use the Hubbard model 
\begin{equation}
\mathcal{H}_C = -t\sum_{i,j,\sigma}a^{\dagger}_{i\sigma} \, a_{j\sigma}
+U\sum_{i}\left(n_{i\uparrow}-\frac12\right)\, \left(n_{i\downarrow}-\frac12\right)
 \, ,
\label{eq1}
\end{equation}
where $a^{\dagger}_{i\sigma}$ ($a_{i\sigma}$) are fermion creation 
(annihilation) operators, and the number operator is given by 
$n_{i\sigma}=a^{\dagger}_{i\sigma}a_{i\sigma}$ with spin $\sigma$ at site 
$i$. The Coulomb-interaction term has been chosen such that the 
grand-canonical ensemble naturally yields a half-filled charge-neutral 
system.

The hopping integral for nearest-neighbors is well known to be $t\approx 
2.7$\,eV~\cite{CastroNeto2009,Yazyev2010}. Since infinite graphene sheets 
are non-magnetic, the local Coulomb interaction $U$ has to be sufficiently 
weak to avoid a bulk magnetic instability, but there is no consensus in 
the literature regarding its precise value. Magnetic resonance 
measurements of neutral soliton states in \textit{trans-polyacetylene} 
have estimated the range $U/t=1.1-1.3$ \cite{Schrieffer1979,Kuroda1987} 
for $sp^2$ carbon systems. First-principle studies based on the local 
density approximation yield $U/t\approx0.9$, while the generalized 
gradient approximation (GGA) leads to $U/t\approx 1.3$ \cite{Bally2000}, 
and a combination of GGA with a constrained random-phase approximation 
gives an onsite Coulomb repulsion $U\approx 9.3\,{\rm eV} \approx 3.4\,t$ 
dressed by some longer-range interactions~\cite{Wehling2011}. In view of 
this uncertainty, we will analyze a range of Coulomb interactions for 
different hexagonal-ZGNF device sizes.

\subsection{Transport formalism}

Electrons need to cross distances of less than 100 atoms when passing
through the nanoflakes that we are going to study here. It is therefore reasonable
to assume that they will not be scattered inside the device and thus 
to treat transport as ballistic. 
Following the Landauer-B\"uttiker formalism \cite{Datta2005}, the 
spin-resolved current thus reads
\begin{equation}\label{eq2}
I_{\sigma} = \frac{e}{h}\int\limits_{-\infty}^{\infty}\mathcal{T}_{\sigma}(\epsilon)\left[f_{L}(\epsilon, T_{L})-f_{R}(\epsilon, T_{R})\right] \, {\rm d}\epsilon\,,
\end{equation}
where $f_{\alpha}(\epsilon, T_{\alpha})$ is the Fermi-Dirac distribution function at the lead $\alpha=L/R$.
$\mathcal{T}_{\sigma}(\epsilon)$ is the spin-resolved transmission for electrons with energy $\epsilon$ evaluated by using the non-equilibrium Green's function  approach,
\begin{equation}\label{eq3}
\mathcal{T}_{\sigma}(\epsilon) = \text{Tr}\Big[\mathbf{\Gamma}_{L}(\epsilon)G^{r}_{\sigma}(\epsilon)\mathbf{\Gamma}_{R}(\epsilon)G^{a}_{\sigma}(\epsilon)\Big]\,,
\end{equation} 
where $\mathbf{\Gamma}_{\alpha}(\epsilon) = 
-i\big[\mathbf{\Sigma}_{\alpha}-\mathbf{\Sigma}^{\dagger}_{\alpha}\big]$ 
is the level broadening caused by the coupling between the lead 
$\alpha=L,R$ and the central region. $G^{r}_{\sigma}$ and $G^{a}_{\sigma}$ 
are the retarded and advanced Green's functions, respectively. We note 
that in the case of the ZGNFs with $N$ sites, both the Green's functions 
and the $\mathbf{\Gamma}_\alpha$ in (\ref{eq3}) are $N \times N$ matrices. 
In general, the functional form of $\mathbf{\Gamma}_{\alpha}(\epsilon)$ 
depends on the details of the hybridization between the electrodes and the 
central region. However, for metals such as gold, the density of states is 
approximately constant near the Fermi energy such that the wide-band limit 
is a good approximation~\cite{Datta2005}. In the 
wide-band limit, the $\mathbf{\Gamma}_{\alpha}(\epsilon)$ are replaced by 
constant matrices 
$\mathbf{\Gamma}_{\alpha}(\epsilon)=\mathbf{\Gamma}_{\alpha}$, where only 
the diagonal entries corresponding to the left or right edge to which the 
respective lead is attached are non-zero. In the following, we choose a 
symmetric coupling such that the non-vanishing matrix elements of 
$\mathbf{\Gamma}_{\alpha}$ all take the same value $\Gamma$. Furthermore, 
we will focus on the value $\Gamma=0.02\,t$ that has been used in a 
previous investigation of the $N=54$ ZGNF \cite{Valli2018}. The Green's 
function needed for the transport computations will be calculated both by 
static mean-field theory (MFT) and a real-space dynamical mean-field 
theory (rDMFT).

\subsection{Static mean-field theory}

Static MFT is a well-established method (see, e.g., chapter 3.1 of 
\cite{Yazyev2010}) and has been employed in previous transport 
computations for graphene nanodevices 
\cite{Zhao2012,Chen2014,Farghadan2018}. The MFT approximation yields an 
effective non-interacting problem which allows the calculation of 
transport coefficients with a standard non-interacting non-equilibrium 
Green's function method. In this approximation, the Hamiltonian of the 
central region (\ref{eq1}) is written as 
$\mathcal{H}_C^{MF}=-t\sum_{i,j,\sigma}a^{\dagger}_{i\sigma}a_{j\sigma}+U\sum_{i\sigma}n_{i\sigma}\langle 
n_{i\bar{\sigma}}\rangle$. Thus, the retarded Green's functions of the 
central region are given by:
\begin{equation}\label{eq4}
G^{r}(\epsilon)= \Big[(\epsilon + i\eta)\mathbb{1}-\mathcal{H}_C^{MF}-\mathbf{\Sigma}_{L}(\epsilon)-\mathbf{\Sigma}_{R}(\epsilon)\Big]^{-1},
\end{equation}
where $\eta$ is an infinitesimal real number, and $\mathbb{1}$ is the 
identity matrix. The expectation values of the occupation number $\langle 
n_{i\bar{\sigma}}\rangle$, acting as mean field, need to be calculated 
selfconsistently. This is done by calculating the local density of states 
$\rho_{i\sigma}(\epsilon)=-\text{Im}~\text{Tr}\big[G^r_{i\sigma}(\epsilon)\big]/\pi$ 
for an electron at site $i$ with spin $\sigma$. The expectation values of 
the occupation operator then reads $\langle n_{i\bar{\sigma}}\rangle= 
\int_{-\infty}^{\epsilon_F}\rho_{i\sigma}(\epsilon) \, {\rm d}\epsilon$. 
By combining this equation with Eq.~(\ref{eq4}), we obtain a set of 
selfconsistency equations that are solved numerically by iteration. This 
selfconsistent solution provides the local spin densities 
$m_i^z=(n_{i\uparrow}-n_{i\downarrow})/2$ on each site.

We note that MFT has been demonstrated to be remarkably accurate for 
static \cite{Feldner2010,*FeldnerE} and in particular dynamic properties 
\cite{Feldner2011} in the semimetallic phase. One of the main shortcomings 
of static MFT is at the quantitative level when approaching the transition 
$U_c$ from the semimetallic phase to the antiferromagnetic insulator. For 
example, the transition point is located at $U_{c}^{\rm MFT}/t \approx 
2.23$ in MFT \cite{Sorella1992} while more sophisticated and accurate 
methods place it in the region $U_c/t \approx 3.8$ 
\cite{Sorella2012,Hassan2013,Assaad2013,Hirschmeier2018}.

\subsection{Dynamical mean-field theory}

\label{sec:DMFT}

The quantitative renormalization upon approaching the transition $U_c$ is 
in turn captured remarkably well by inclusion of local charge fluctuations 
\cite{Marcin2019} in the framework of a single-site dynamical mean-field 
theory \cite{Georges1996}. Thus, in order to improve quantitative accuracy 
with respect to $U$, here we also employ real-space dynamical mean field 
theory (rDMFT) to obtain a magnetic solution of the graphene flake.

rDMFT is a non-perturbative approach which maps the lattice Hamiltonian 
Eq.~(\ref{eq1}) onto a set of quantum impurity models~\cite{Georges1996}. 
This mapping is performed by calculating the local Green's functions of 
all lattice sites
\begin{equation}
G^r_i(z)=(z-H_0-\mathbf{\Sigma}^r(z))^{-1}_{ii}\,, 
\label{eq:rDMFT}
\end{equation}
where $H_0$ is the single-particle part of Eq.~(\ref{eq1}) and 
$\mathbf{\Sigma}^r(z)$ the retarded self-energy matrix. The index $i$ 
corresponds to a lattice site. The local Green's function for lattice site 
$i$ can be written as
\begin{equation} 
G^r_i(z)=\frac{1}{z-\Delta_i(z)-\Sigma^r_i(z)} \, , \label{eq:rDMFT2} 
\end{equation}
where $\Sigma^r_i(z)$ is the retarded local self energy of this 
lattice site. This equation defines a hybridization function $\Delta_i(z)$ 
for each lattice site, which can be used to set up a single-impurity 
Anderson model. We use the numerical renormalization group (NRG) to solve 
these impurity models and obtain the self energies for all lattice 
sites~\cite{Wilson1975,Krishna1980,Bulla2008}. The obtained self energies 
are used to calculate new local Green's functions according to 
Eq.~(\ref{eq:rDMFT}). Equations (\ref{eq:rDMFT}) and (\ref{eq:rDMFT2}), 
together with the NRG, are iterated until selfconsistency is reached. In 
this way, rDMFT can be used to calculate magnetic solutions of finite 
clusters~\cite{Robert2014,Robert2015}.

Using NRG for finding the self energies gives us immediate access to 
real-frequency Green's functions and self energies with high accuracy 
around the Fermi energy~\cite{Robert2006}. After selfconsistency is 
reached, we can use these Green's functions and self energies to evaluate 
the transmission coefficient, Eq.~(\ref{eq3}). We use the Dyson equation 
$G(\epsilon)^{-1}=G_0(\epsilon)^{-1}-\mathbf{\Sigma}_{L}(\epsilon)-\mathbf{\Sigma}_{R}(\epsilon)-\mathbf{\Sigma}(\epsilon)$, 
where $G_0(\epsilon)^{-1}$ is the bare Green's function of the central 
region, and $\mathbf{\Sigma}(\epsilon)$ is the self energy obtained by 
NRG.

We note that, thanks to a logarithmic energy discretization, NRG yields 
high frequency resolution close to the Fermi energy, but lower resolution 
at higher energies, see Ref.~\cite{Bulla2008} for details.

\section{Results}
\label{secIII}

In this chapter we present our results for the transport properties of 
hexagonal ZGNFs. However, before we do so, we briefly revisit their 
magnetic properties.

\subsection{Edge magnetism of hexagonal ZGNFs}

\begin{figure}
 \centering
\includegraphics[width=1\columnwidth]{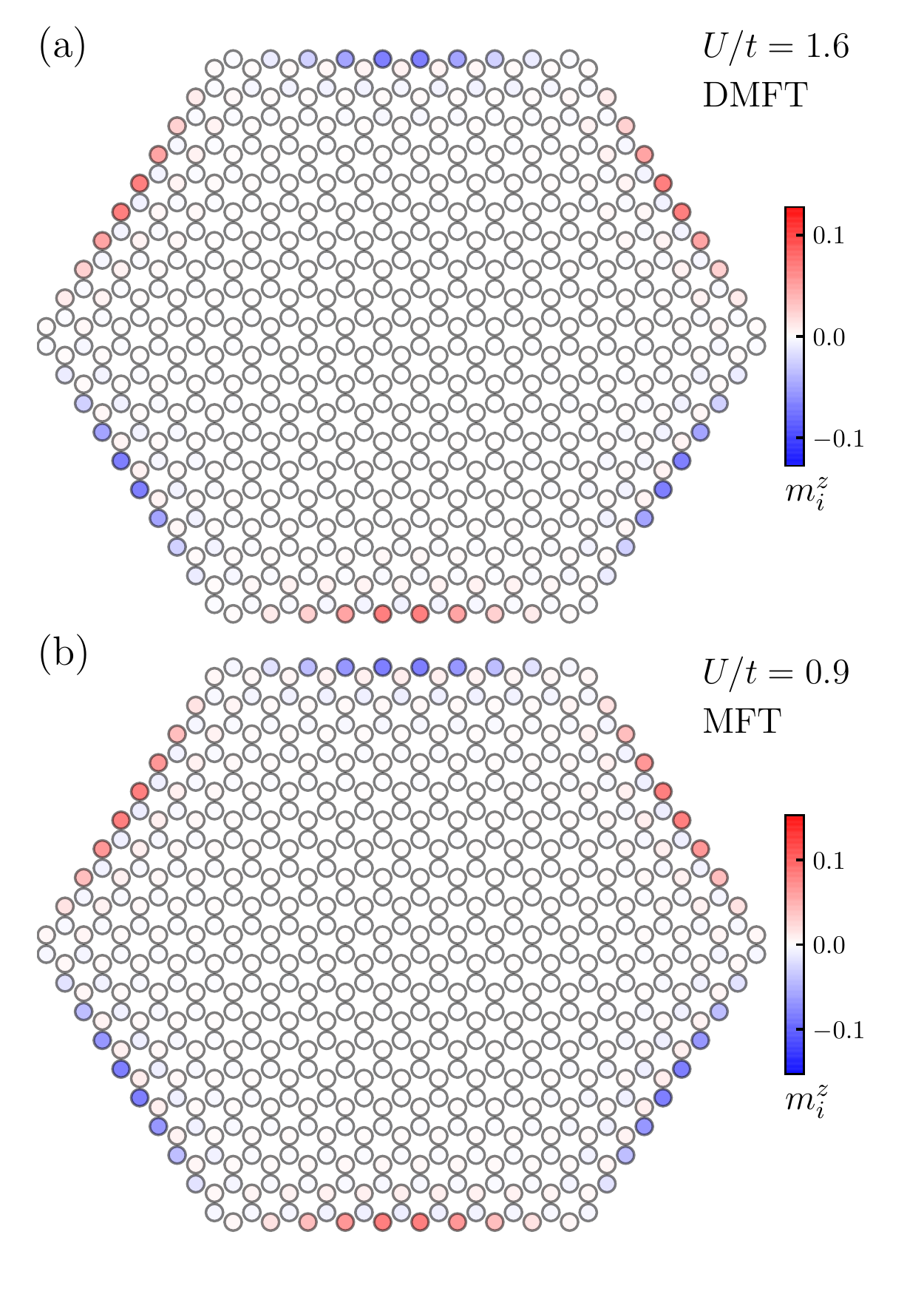}
\caption{Spatial magnetization profile for a flake of size $N=600$ based on (a) DMFT ($U/t=1.6$) and (b) MFT ($U/t=0.9$).}
\label{fig2b}
\end{figure}

\begin{figure}
 \centering
\includegraphics[width=1\columnwidth]{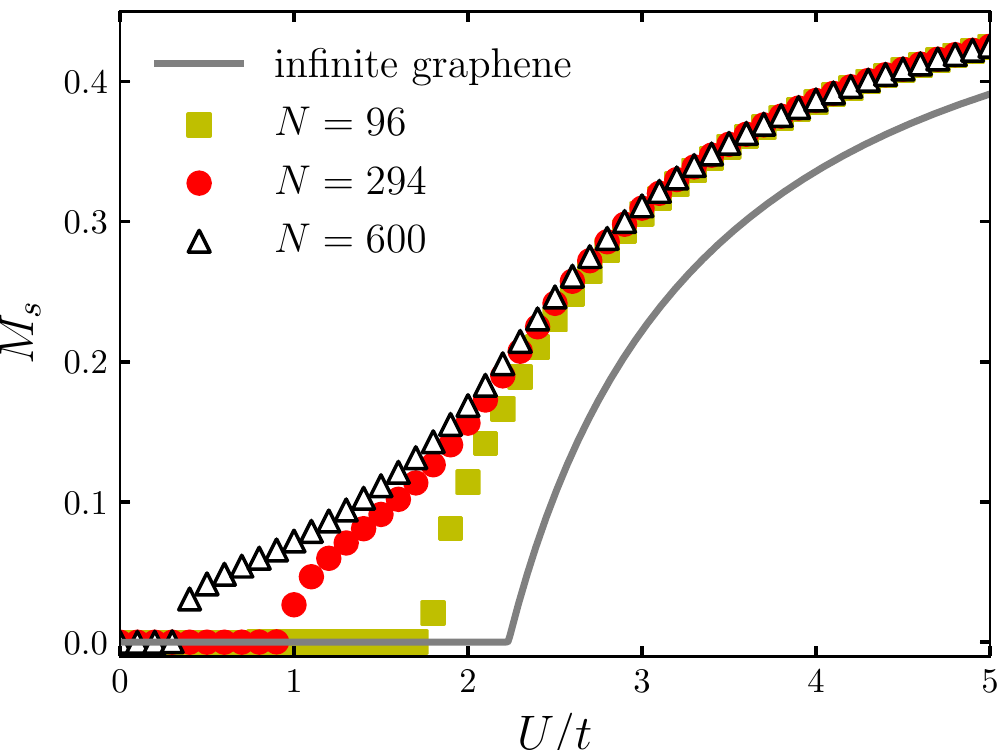}
\caption{Average staggered magnetization $M_s$ of edge sites versus $U/t$ 
based on MFT. Three different sizes are shown by symbols; the gray line shows 
the staggered magnetization for an infinite graphene sheet. }
\label{fig2}
\end{figure}

Hexagonal ZGNFs constitute a well-studied 
example~\cite{Fernandez2007,Bhowmick2008,Viana2009,Feldner2010,*FeldnerE,Roy2014,Valli2016} 
of magnetism at zigzag edges. Here we illustrate this with the flake shown 
in Fig.~\ref{fig2b}. This flake contains $N=600$ carbon atoms 
and is thus bigger than most examples previously studied in the 
literature~\cite{Fernandez2007,Bhowmick2008,Viana2009,Feldner2010,*FeldnerE,Valli2016}. 
The localized electronic states at the edges of the ZGNFs lead to a 
nonuniform spatial profile of the local magnetic moments $\langle m_i^z 
\rangle$ which are also mainly localized on the outermost sites. This is illustrated 
by the color scale in Fig.~\ref{fig2b} that shows the magnetic 
moments for the hexagonal ZGNF with $N=600$ sites. One observes that spins 
align ferromagnetically at each edge already for a value of
$U/t=1.6$ ($U/t=0.9$), well  below the bulk (D)MFT critical value 
$U^{\rm DMFT}_c/t = 3.5 \ldots 3.7$~\cite{Marcin2019}
($U^{\rm MFT}_c/t\approx2.23$~\cite{Sorella1992}). This ferromagnetic alignment 
corresponds to all carbon atoms on each edge belonging to the same 
sublattice. Adjacent edges are separated by an armchair defect, belong to 
different sublattices, and thus have opposite magnetization.
We note that the magnetic pattern in Fig.~\ref{fig2b} is very similar
for DMFT and MFT, just the value of the Coulomb interaction is renormalized,
as expected from a previous investigation of infinite graphene sheets \cite{Marcin2019}.

Given that the magnetic moment is localized at the edge, its relative
contribution to the total staggered magnetization of the flake
will vanish in the thermodynamic limit below the bulk critical $U_c$.
Therefore, we characterize this phenomenon by the average staggered magnetization
\begin{equation}
M_s=\frac{1}{N_{\rm edge}}\sum_{i\in {\rm edge}}^{N_{\rm edge}}
\zeta_i\, \langle m_i^z \rangle
\label{wq:Ms}
\end{equation}
at the edge where $\zeta_i =1$ ($-1$) for $i$ in the $A$ ($B$) sublattice.
Figure \ref{fig2} shows MFT results for $M_s$, and
one observes that the critical point where the edge of the 
hexagonal ZGNF becomes magnetic decreases considerably with increasing 
size. At the biggest size $N=600$ studied here, the critical point shifts 
down to the range $\left.U_c^{\rm MFT}/t\right|_{N=600} \approx 
(0.3-0.4)$, which is far below the critical point of bulk graphene 
(compare the gray curve in Fig.~\ref{fig2}). This critical point is also 
significantly below the lower bounds for the on-site Coulomb repulsion 
$U/t \approx 1$ in graphene~\cite{Schrieffer1979,Kuroda1987,Bally2000}, 
already mentioned in Sec.~\ref{sec:Model}.
Figure \ref{fig2b}(b) shows that magnetic edges appear for the $N=600$ ZGNF
already for $U/t \le 1.6$ within real-space DMFT. Further calculations
(not shown here) show a 
shift of the critical point within rDMFT to a renormalized range $\left.U_c^{\rm 
DMFT}/t\right|_{N=600}\approx (1.2-1.5)$ for the $N=600$ system. This range is again well below 
the corresponding estimate for the bulk critical point $3.5 \lesssim 
U^{\rm DMFT}_c/t \lesssim 3.7$~\cite{Marcin2019}. However, this presumably 
more realistic estimate now becomes comparable to the estimates of the 
Coulomb interaction. One nevertheless concludes that the magnetic 
instability at the zigzag edges can be observed in a range of on-site 
Coulomb interactions that are realistic for graphene, and that the balance 
could be shifted to even smaller $\left.U_c\right|_N$ by going to larger 
sizes $N>600$.

\begin{table}[t!]
\centering
\begin{tabular}{c|c}
$N$   & $\Delta_{\rm sp}/t$ \\ \hline
$54$  & $0.342041$ \\
$96$  & $0.229448$ \\
$294$ & $0.072607$ \\
$600$ & $0.021013$ \\
\end{tabular}
\caption{Single-particle gap $\Delta_{\rm sp}$
for non-interacting electrons ($U=0$) on a hexagonal
ZGNF as a function of its size $N$.
\label{tab:dotGapN}
}
\end{table}

The magnetic state located at the edge of a ZGNF is closely related to the 
single-particle level closest to the Fermi energy in the regime 
$\left.U_c\right|_N \le U < U_c$. This level is separated from the Fermi 
energy by a single-particle gap $\Delta_{\rm sp}$. Some values of this gap 
for the non-interacting ($U=0$) system are quoted in Table 
\ref{tab:dotGapN}. One observes that this single-particle gap decreases 
rapidly with increasing size of the flake $N$, thus giving rise to the 
reduction of $\left.U_c\right|_N$ with growing $N$ observed in 
Fig.~\ref{fig2} and discussed above, see also chapter 3.3 of 
Ref.~\cite{phdThu}. We note that this reduction of the critical 
$\left.U_c\right|_{N=54}$ relative to the bulk value of $U_c$ is almost 
absent for the $N=54$ flake studied in 
Refs.~\cite{Valli2016,Valli2018,Valli2019}.

\subsection{Spin-resolved transmission in hexagonal ZGNFs}

\label{sec:Trans}

\begin{figure}
 \centering
\includegraphics[width=1\columnwidth]{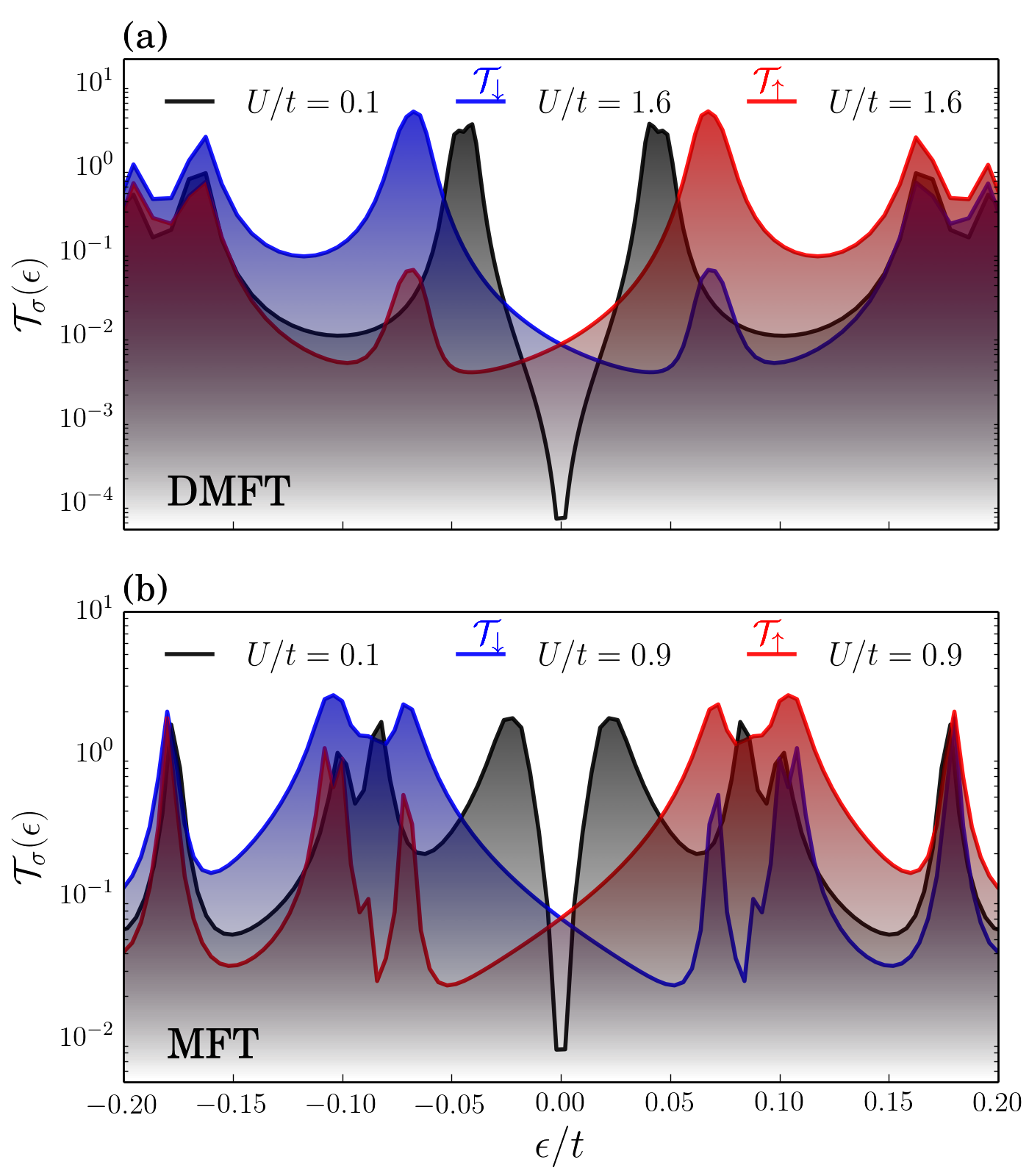}
\caption{Spin-resolved transmission coefficients $\mathcal{T}_{\sigma}$ 
calculated by (a) DMFT and (b) MFT as function of $\epsilon/t$ for the 
\textit{meta} configuration of a hexagonal ZGNF with size $N=600$.
The coupling to the leads was  taken to be $\Gamma = 0.02\,t$.
In both 
panels, the gray curve is for a Coulomb interaction $U/t = 0.1$ that
is below the edge critical value $\left.U_c\right|_{N=600}$. The values
$U/t=1.6$ and $0.9$ in panel (a) and (b), respectively, lie between $\left.U_c\right|_{N=600}$
and the bulk critical value $U_c$. In the latter case, the red and blue
curves are the transmission functions ${\cal T}_\uparrow$ and ${\cal T}_\downarrow$
in the two different spin channels.
 }
\label{fig3}
\end{figure}

Next, we use MFT and rDMFT to compute the transmission functions of the 
flakes. As mentioned in Sec.~\ref{secII}, rDMFT is more expensive, but 
expected to yield a quantitatively more accurate account of the Coulomb 
integration $U$ in the Hubbard model (\ref{eq1}).

Figures \ref{fig3}(a) and (b) show the DMFT and MFT results of the 
spin-resolved transmission $\mathcal{T}_{\sigma}(\epsilon)$ in the 
\textit{meta} configuration as function of the energy of the electrons, 
respectively. We compare results for two values of the Coulomb 
interaction; one interaction strength is larger than the critical value 
$\left.U_c\right|_N$ for this flake size, the other is smaller. In the 
former case, the transmission functions shown in Fig.~\ref{fig3} 
correspond to the magnetic states shown previously in Fig.~\ref{fig2b}. 
The main difference between MFT and DMFT is the renormalization by charge 
fluctuations from $U/t=0.9$ in MFT to $U/t=1.6$ in DMFT. The broadening 
observed in Fig.~\ref{fig3} is due firstly to the coupling to the leads, 
$\Gamma = 0.02\,t$. For the MFT computation we chose $\eta = 10^{-6}\,t 
\ll \Gamma$ such that its effect is negligible. However, in the framework 
of DMFT there is an additional non-uniform broadening in particular at 
higher energies $\left|\epsilon\right|$, owing to the logarithmic 
frequency resolution of the NRG impurity solver (see discussion at the end 
of Sec.~\ref{sec:DMFT}).

According to Fig.~\ref{fig2b}, the \textit{meta} configuration corresponds 
a situation where the leads in Fig.~\ref{fig1} are attached to edges with 
the same magnetic polarization. Thus, for sufficiently large Coulomb 
interaction, the hexagonal ZGNF is magnetically ordered and the spin 
degeneracy is lifted resulting in a distinct spin-dependent transmission 
$\mathcal{T}_\sigma(\epsilon)$. While the asymmetric transmission spectra 
for each spin direction $\mathcal{T}_\sigma(\epsilon)\neq 
\mathcal{T}_{\sigma}(-\epsilon)$ gives rise to the spin Seebeck effect 
reported in recent works~\cite{Sierra2018,Tang2018}, the particle-hole 
symmetry of the transmission coefficient 
$\mathcal{T}_\sigma(\epsilon)=\mathcal{T}_{\bar{\sigma}}(-\epsilon)$ under 
spin inversion leads to a pure spin current, as is discussed later. 
Figure~\ref{fig3} also shows that in the non-magnetic state an 
anti-resonance (a deep valley) appears at the Fermi energy in the 
\textit{meta} configuration. We have checked other possible configurations 
of the leads (results not shown here), where the leads would be attached 
to edges of different spin polarization in Fig.~\ref{fig2b}, that are 
known as \textit{ortho} and \textit{para} configuration~\cite{Borges2017}, 
and found no spin-dependent transmission. Thus, in the following, we 
present results only for the \textit{meta} configuration.

The lowest peaks in Fig.~\ref{fig3} appear at $\left|\epsilon\right| 
\approx 0.07\,t$ for $U/t = 1.6$ and $0.9$ in DMFT and static MFT, 
respectively. This corresponds to a single-particle gap of approximately 
$0.07\,t$, that is filled in by the coupling to the leads, $\Gamma = 
0.02\,t$. Note that the single-particle gap $\Delta_{\rm sp}$ is already 
enhanced significantly as compared to its non-interacting value even by 
these moderate values of the Coulomb interaction, see Table 
\ref{tab:dotGapN}. The temperature scale corresponding to the values of 
$\Delta_{\rm sp} \approx 0.07\,t$ is on the order of 2200\,K and thus 
significantly higher than the temperatures that we will study below. 
Indeed, we have checked that a temperature of $T/t=1/100$ has no visible 
effect on the MFT results for the magnetic state shown in 
Fig.~\ref{fig2b}(b) and should thus also be negligible for the 
transmission functions shown in Fig.~\ref{fig3}. This justifies using 
these zero-temperature transmission functions to compute 
finite-temperature transport properties, as we will do in the following.

\begin{figure}[t!]
\centering\includegraphics[width=1.0\columnwidth]{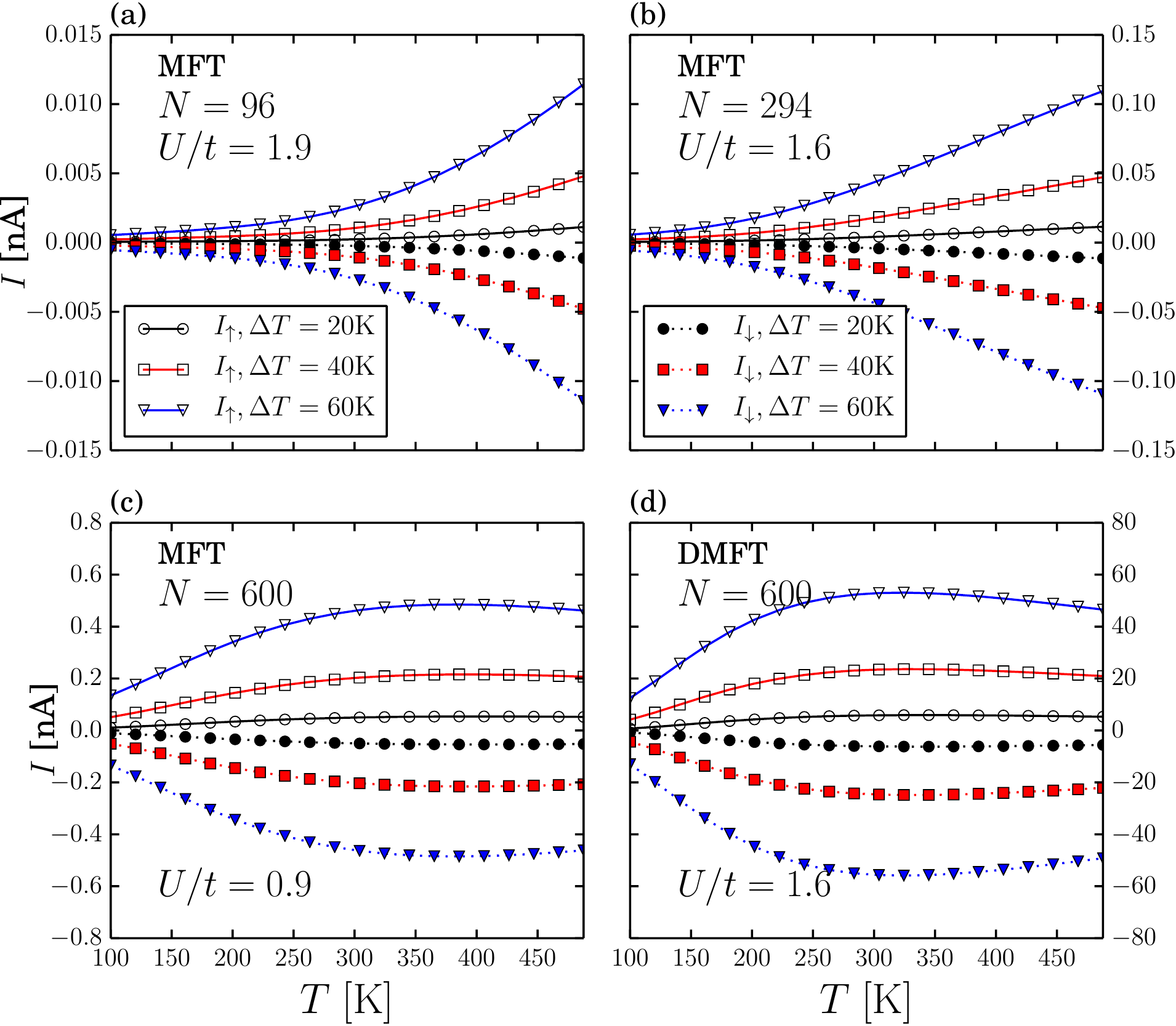}\\[2mm]
\includegraphics[width=.48\columnwidth]{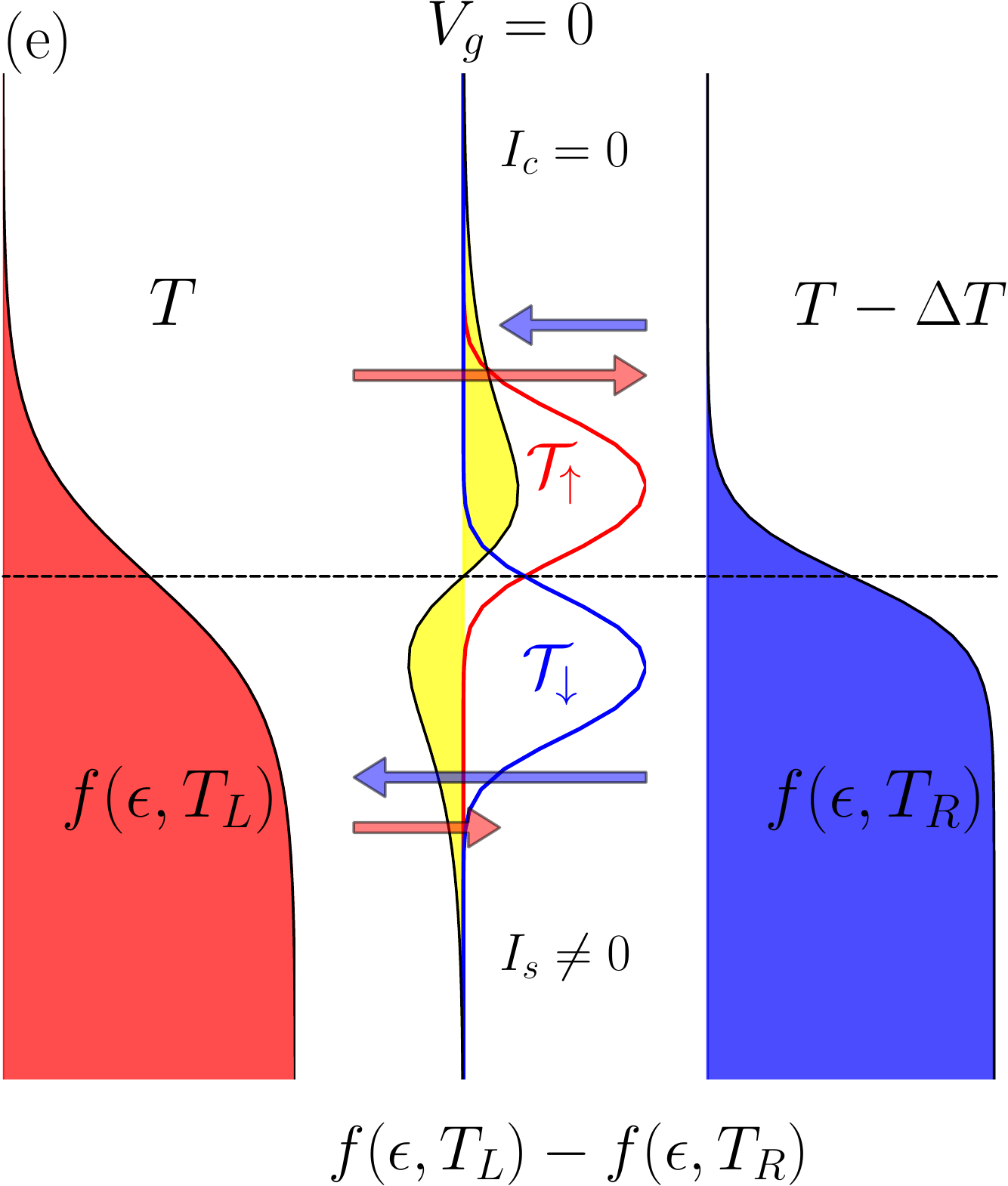}\hfill%
\includegraphics[width=.48\columnwidth]{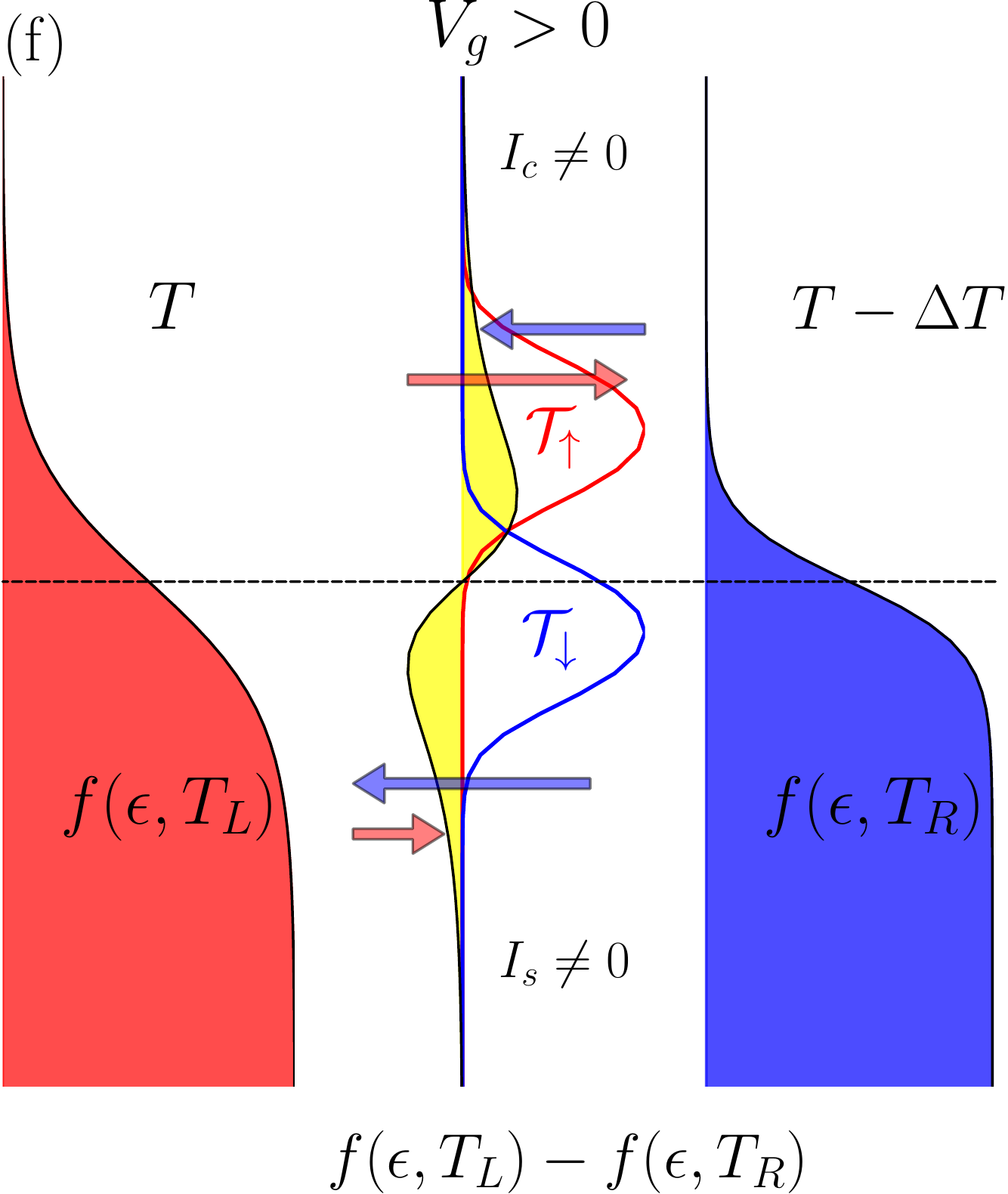}
\caption{Spin-resolved current versus temperature $T$ for several 
temperature differences $\Delta T$ and flakes of size (a) $N=96$, (b) 
$N=294$, and (c,d) $N=600$. Note the different vertical scales of the 
different panels. The coupling to the leads has been chosen as 
$\Gamma=0.02\,t$. Panels (a--c) are based on MFT, panel (d) is based on 
DMFT. The value of $U/t$ is adjusted according to Fig.~\ref{fig2} (compare 
also the discussion in the main text). Panels (e) and (f) show the 
Fermi-Dirac distributions of the hot and cold leads, respectively, for a 
gate voltage $V_g=0$ (e) and $V_g>0$ (f); the spin-resolved transmission 
functions are illustrated schematically in the middle between the two 
Fermi functions. The resulting currents are indicated separately for 
spin-up (red arrows) and spin-down electrons (blue arrows), as well as for 
electrons with energies higher (upper set of arrows) and lower (lower set 
of arrows) than the equilibrium chemical potential (black dotted 
horizontal line). The yellow filled curve represents the difference of the 
two Fermi-Dirac distributions.
}
\label{fig4}
\end{figure}

\subsection{Spin-resolved current in hexagonal ZGNFs}
 
Having calculated the spin-resolved transmission functions, we calculate 
the current with Eq.~(\ref{eq2}). Figure~\ref{fig4} shows the spin-up and 
spin-down currents as a function of the temperature $T = T_L$ for several 
fixed temperature differences, $\Delta T = T_L - T_R$, between the leads. 
As can be seen, spin-up ($I_{\uparrow}$) and spin-down currents 
($I_{\downarrow}$) are of different sign but identical magnitude. The 
values of $U/t$ are chosen in each case such as to yield magnetic edges 
without rendering the bulk of the flake magnetic. We recall that MFT has a 
higher frequency resolution, but we expect DMFT to provide a more accurate 
description of Coulomb interactions, and that the values of $U/t=0.9$ and 
$1.6$ in panels (c) and (d) of Fig.~\ref{fig4} were chosen such as to 
account for the renormalization by charge fluctuations. After this 
renormalization, the shape of the curves becomes very similar, as is to be 
expected given the similarity of the underlying transmission functions 
already seen in Fig.~\ref{fig3}.

We note that the absolute value of the currents in Fig.~\ref{fig4} depends 
on the choice of $\Gamma$, not only explicitly in Eq.~(\ref{eq3}), but 
also via the Green's functions, see Eq.~(\ref{eq4}) and recall that 
$\mathbf{\Sigma}_\alpha \propto i\,\Gamma$. However, we have checked that this only 
affects the absolute value of $I$ with a power of $\Gamma$ that appears to 
be actually lower than 2 while the qualitative temperature dependence of 
the currents is not very sensitive to the precise value of $\Gamma$. In 
MFT, we chose $\eta=10^{-6}\,t$ such that this additional broadening 
becomes irrelevant. However, we recall that the NRG impurity solver for 
DMFT only yields a logarithmic frequency resolution that cannot be 
eliminated. The experience with the static MFT indicates that this may 
lead to an overestimatation of the absolute value of the currents in 
Fig.~\ref{fig4}(d) by one to two orders of magnitude, but we still expect 
the qualitative behavior as a function of temperature to be not very 
sensitive to the limited energy resolution also in this case.

One observes that the spin currents increase rapidly with temperature for 
the larger systems, yielding currents that in MFT increase from a few pA 
to $I \approx 0.5$\,nA around room temperature and for a temperature 
difference $\Delta T= 60$\,K, as $N$ increases from $96$ to $600$. 
Remarkably, the $N=600$ flake not only yields the largest current, but 
also exhibits a maximum of the spin current around room temperature, both 
within MFT and DMFT, compare Fig.~\ref{fig4}(c,d). The currents set on 
around a threshold temperature $T_{\text{th},\sigma}$ that is equal for up 
and down spin currents. While $T_{\text{th},\sigma}$ is not very sensitive 
to the temperature difference $\Delta T$, it exhibits a strong dependence 
on the size of the hexagonal ZGNF as $T_{\text{th},\sigma}$ decreases 
significantly with increasing size. For ZGNRs, the size dependence of the 
threshold temperatures $T_{\text{th},\sigma}$ has been investigated for 
width 6 and 14, yielding only a slight decrease~\cite{Zeng2011}. This weak 
dependence of $T_{\text{th},\sigma}$ on the width of ZGNRs can be 
understood since such ribbons have zero-energy states for $U=0$ that arise 
from the infinitely long zigzag edges for any width 
\cite{Fujita96,Hikihara2003,Son2006,Fernandez2008,Yazyev2010} whereas 
non-interacting ZGNFs have a finite-size gap in their single-particle 
spectrum that decreases with increasing $N$ as shown in Table \ref{tab:dotGapN}.
However, in contrast to 
hexagonal ZGNFs, ZGNR-based devices show $T_{\text{th},\uparrow}\neq 
T_{\text{th},\downarrow}$. Furthermore, these spin currents saturate 
around room temperature for the $N=600$ flake. The saturation current for 
ZGNR-based spin caloritronics devices is not discussed in the literature, 
but one can expect it to occur above room temperature, at least for the 
width-6 and -14 ZGNRs studied in Ref.~\cite{Zeng2011}.

The reduction of $T_{\text{th},\sigma}$ can be understood in terms of the 
overlap of the transmission functions $\mathcal{T}_{\sigma}(\epsilon)$ 
with the difference of the Fermi functions $\big[f_{L}(\epsilon, 
T_{L})-f_{R}(\epsilon, T_{R})\big]$ in Eq.~(\ref{eq2}). Indeed, larger 
hexagonal ZGNFs have smaller single-particle gaps and thus provide for 
more transport channels close to the Fermi energy. Thus, one can expect a 
bigger contribution from the integrand in Eq.~(\ref{eq2}). This larger 
overlap leads firstly to a larger current. Secondly, the reduced energy 
difference of the transmission peaks and the Fermi energy gives rise to a 
lower threshold temperature $T_{\text{th},\sigma}$. The relatively low 
$T_{\text{th},\sigma}$ for the flake size $N=600$ in Fig.~\ref{fig4}(c) 
and (d) can be understood by the energy difference between the highest 
occupied and lowest unoccupied states shown by the positions of first 
peaks of the transmission spectrum in Fig.~\ref{fig3}(a) and (b), compare 
also the related discussion at the end of Sec.~\ref{sec:Trans}, although 
$T_{\text{th},\sigma}$ is evidently smaller than this bare electronic 
energy scale.

For a more detailed understanding of the underlying mechanism that 
generates these currents in the device, we analyze the dependence of the 
spin current on the Fermi-Dirac distribution functions of the leads and 
the transmission function in Eq.~(\ref{eq2}). Fig.~\ref{fig4}(e) 
schematically shows the transmission function of the junctions and the 
Fermi-Dirac distributions of the hot (left) and cold (right) leads. Since 
$\mathcal{T}_{\uparrow}(\epsilon)$ is larger above the Fermi energy, 
charge carriers with high energy flow from the hot lead to the cold lead, 
giving rise to a positive spin-up current from the hot to the cold lead. 
Conversely, $\mathcal{T}_{\downarrow}(\epsilon)$ is much larger below the 
Fermi energy (Fig.~\ref{fig4}(e)) such that the spin-down current flows in 
the opposite direction. In fact, thanks to the symmetry under the combined 
particle-hole transformation and spin inversion, the transmission 
functions satisfy 
$\mathcal{T}_\sigma(\epsilon)=\mathcal{T}_{\bar{\sigma}}(-\epsilon)$, 
where $\bar{\sigma}$ is the spin projection opposite to $\sigma$. This 
implies that the total charge current, 
$I_c=I_{\uparrow}+I_{\downarrow}=0$, is exactly zero and the device 
exhibits a pure spin current, $I_s=I_{\uparrow}-I_{\downarrow}\neq0$.

Returning to the case of ZGNRs, we note that the two opposite edges are 
individually ferromagnetic in the ground state, but are correlated 
antiferromagnetically (see 
Refs.~\cite{Fujita96,Wakabayashi98,Son2006,Fernandez2008,Yazyev2010,Feldner2011,Magda2014} and 
references therein). However, a ZGNR-based spin-caloritronics device 
requires ferromagnetic ZGNRs, at least as long as leads extend over the 
full width of the ribbon. Consequently, a strong external magnetic 
field~\cite{Son2006}, doping, or width engineering~\cite{Chen2017} are 
needed to stabilize a ferromagnetic state. From this point of view, the 
hexagonal-ZGNF junction proposed here has a big advantage over the 
ZGNRs~\cite{Zeng2011,Zhai2019,LV2019}, and also armchair silicene 
nanoribbons~\cite{Tan2018} for spin-caloritronic applications.

\subsection{Effect of the Coulomb interaction}

\begin{figure}[t!]
\includegraphics[width=\columnwidth]{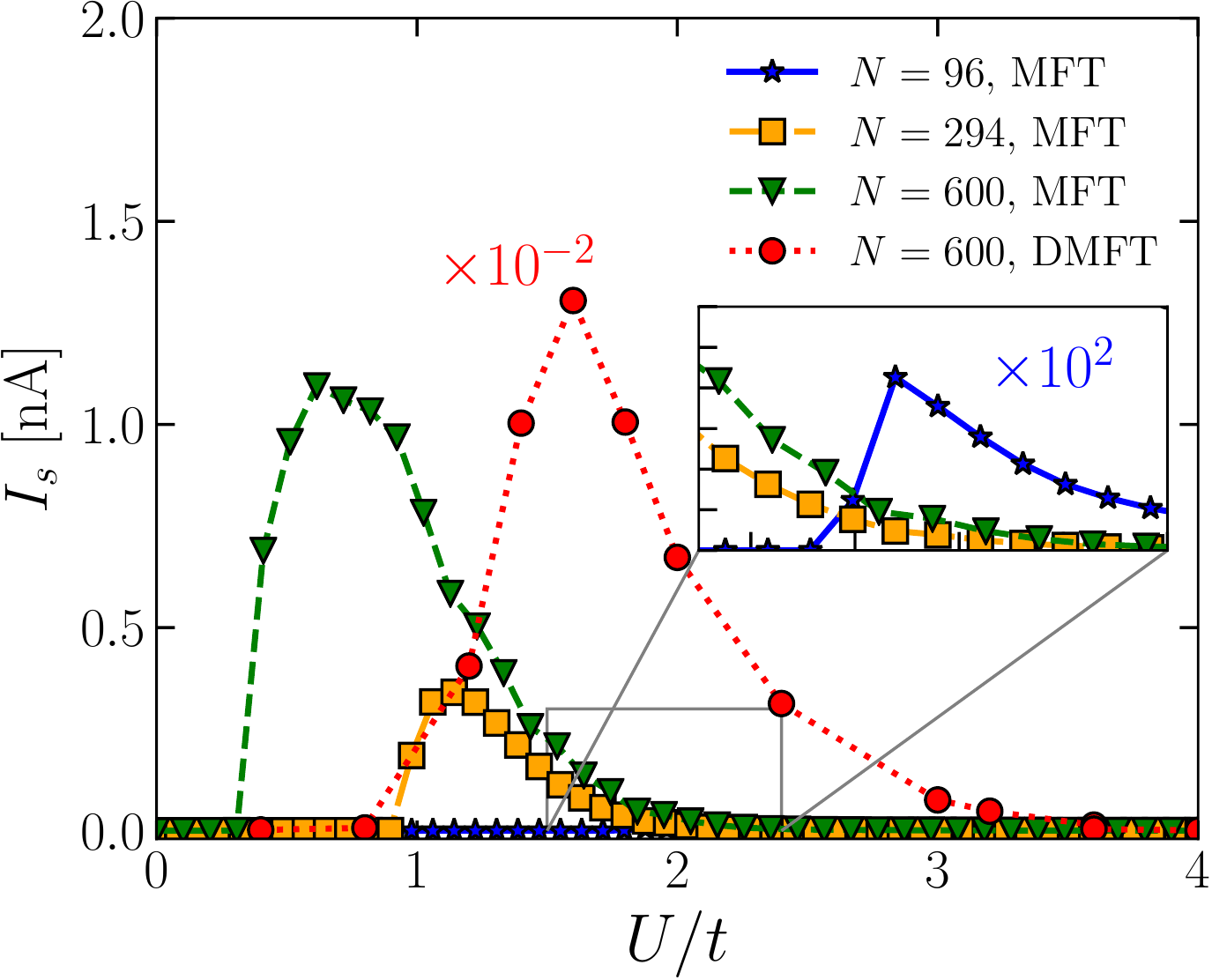}
\caption{Dependence of the spin current $I_s=I_\uparrow - I_\downarrow$ on 
the Coulomb repulsion $U/t$ and flake size at room temperature 
$T=300\,{\rm K}$ for a fixed temperature difference $\Delta T=60\,{\rm K}$ 
and coupling to the leads $\Gamma=0.02\,t$. The inset magnifies a portion 
of the current where the data for the flake with $N=96$ sites is multiplied by 
a factor $100$ for better visibility; the DMFT data for $N=600$ has been 
divided by $100$.
}
\label{fig5}
\end{figure} 

Since the exact value of the Coulomb interaction in hexagonal ZGNFs is 
unknown, we show the spin current $I_s=I_\uparrow - I_\downarrow$ as a 
function of on-site Coulomb repulsion $U/t$ in Fig.~\ref{fig5} for three 
different flake sizes where for $N=600$ we again compare MFT and DMFT 
results. We note that $I_\downarrow = -I_\uparrow$ since we are still at 
charge neutrality (half filling) such that $I_s=2\,I_\uparrow$. The DMFT 
results and MFT results in Fig.~\ref{fig5} for the $N=600$ flake differ by 
about a factor $100$, as we have already seen in the context of 
Fig.~\ref{fig4}(c,d). We recall that this may be due to an overestimation 
of the absolute value of the current within DMFT owing to the logarithmic 
frequency resolution.

A first observation from Fig.~\ref{fig5} is that the spin current vanishes 
for interaction strengths smaller than the critical point, 
$\left.U_c\right|_N$. This is expected because a magnetically ordered edge 
is needed to establish a spin current. As was already observed in 
Fig.~\ref{fig2}, this critical $\left.U_c\right|_N$ is size-dependent 
which is also visible in the spin current. For $U>\left.U_c\right|_N$, the 
edge becomes magnetic which results in a strong increase of the spin 
current. Figure \ref{fig5} shows two curves for the $N=600$ ZGNF. These 
two curves have a similar shape; the DMFT curve is just shifted to larger 
values of $U/t$ with respect to the MFT curve owing to the already 
mentioned renormalization by charge fluctuations that are ignored in MFT.

Figure~\ref{fig5} also shows again that the maximal attainable spin 
current grows rapidly with increasing flake size; for the $N=96$ flake we 
need to multiply it by $10^2$ to render it even visible on the scale of 
the other curves. For flakes with $N=600$ sites and the parameters 
$T=300\,{\rm K}$, $\Delta T=60\,{\rm K}$, the spin current reaches a 
maximal value $I_s\sim1\,\text{nA}$ within MFT. However, a further 
increase of the Coulomb interaction leads to a decrease of the spin 
current until it vanishes at a specific point which is almost 
size-independent. This interaction strength is close to the bulk critical 
point of the honeycomb Hubbard model, that is given by $U_{c}^{\rm MFT}/t 
\approx 2.23$ in MFT \cite{Sorella1992} and $U^{\rm DMFT}_c/t = 3.5 \ldots 
3.7$~\cite{Marcin2019}, respectively. Beyond this bulk critical point, the 
system becomes an antiferromagnetic Mott insulator and all currents are 
blocked in the device.

We mention that one can see qualitatively the same behavior for other 
temperatures $(T,\Delta T)$.

\subsection{Effect of a gate voltage} 

\begin{figure}[t!]
\centering
\includegraphics[width=1\columnwidth]{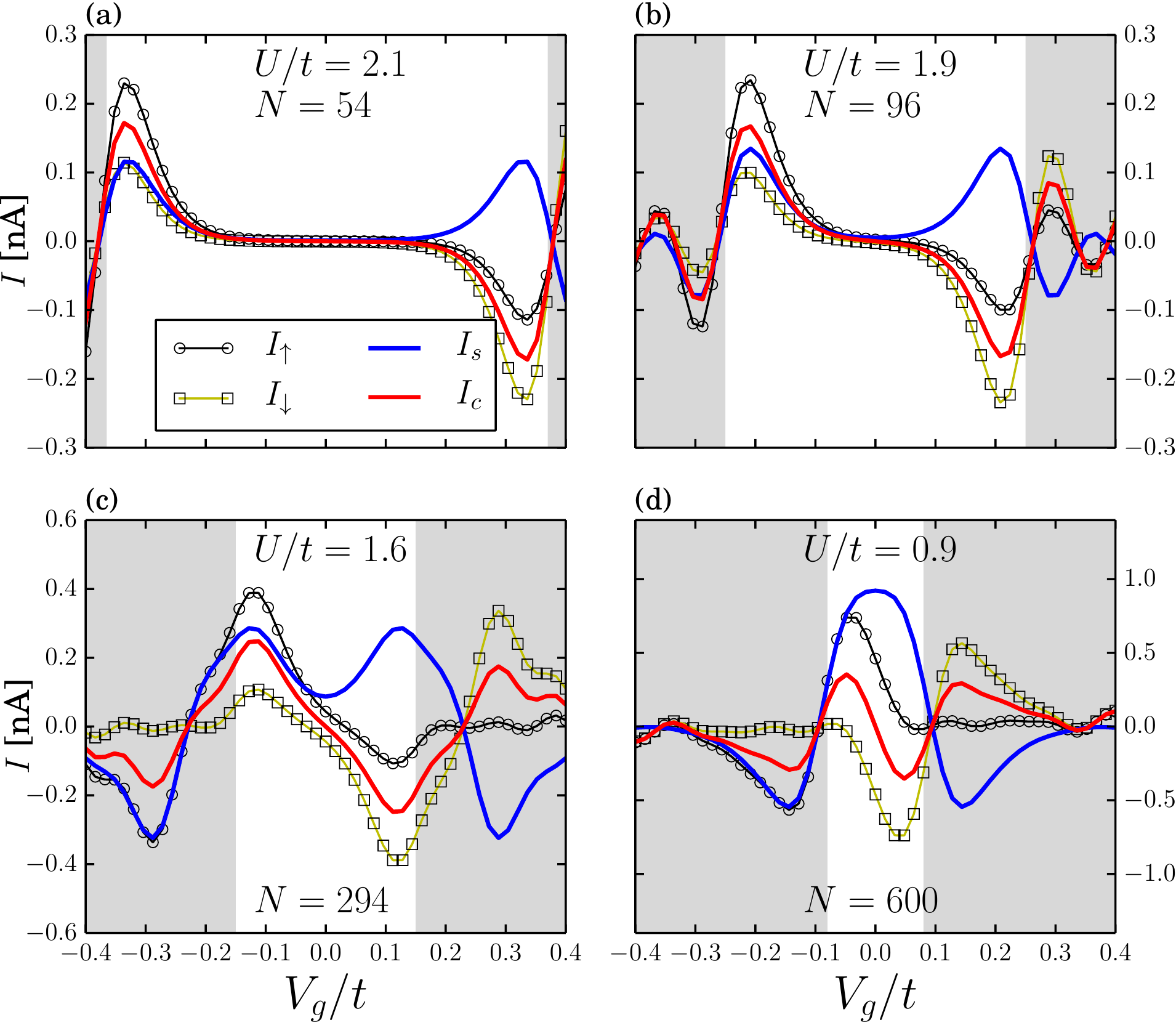}
\caption{Spin-resolved current versus gate voltage $V_g/t$ for flakes with 
sizes (a) $N=54$, (b) $N=96$, (c) $N=294$, and (d) $N=600$ based on MFT. 
The temperature is $T=300\,{\rm K}$, the temperature difference is $\Delta 
T=60\,{\rm K}$, and the coupling to the leads $\Gamma = 0.02\,t$. In each 
case, we have chosen a value of $U$ such that the edge is magnetic but the 
bulk of the flake remains non-magnetic. Note the different vertical scales 
of the panels.
}
\label{fig6}
\end{figure}

Finally, we discuss the effect of a gate voltage $V_g$. We assume that a 
positive $V_g$ raises the Fermi level of the flake relative to the leads, 
as sketched in Figure~\ref{fig4}(f), but otherwise has no effect such that 
we can use the transmission functions $\mathcal{T}_{\sigma}(\epsilon+V_g)$ 
that have been obtained at half filling, just shifting their argument by 
$V_g$ in Eq.~(\ref{eq2}). We note that when the gate voltage exceeds the 
single-particle gap, electrons are doped into the system which might 
change the magnetic state of the flake. We leave a detailed investigation 
of the interplay of doping and the magnetic state to future studies and 
focus our discussion on values of the gate voltage inside the gap.

Figure~\ref{fig6} shows the dependence of the spin currents on the gate 
voltage at room temperature with a fixed temperature difference $\Delta 
T=60\,{\rm K}$. In each case, we have chosen a value of $U$ such that 
$\left.U_c^{\rm MFT}\right|_N < U < U_c^{\rm MFT}$. Here we also included 
an example for $N=54$ in Fig.~\ref{fig6}(a) where the aforementioned 
condition forces us to choose $U/t=2.1$, very close to the bulk transition 
$U^{\rm MFT}_c/t\approx2.23$~\cite{Sorella1992}. Regions where the gate 
voltage exceeds the single-particle gap are shaded in gray in 
Fig.~\ref{fig6} and will be excluded from further discussion.

Although the symmetry of the spin-up and spin-down transmission 
coefficients is preserved, the shift of the chemical potential 
(\textit{i.e.}, the Fermi level) on the flake results in a situation where 
the spin-down current is larger than the spin-up current for $V_g>0$ and vice 
versa for $V_g<0$. Indeed, the spin-resolved currents are antisymmetric 
under spin inversion $I_{\sigma}(V_g)=-I_{\bar{\sigma}}(-V_g)$ which leads 
to a spin current $I_s$ that is even under sign inversion of $V_g$ and a 
charge current $I_c$ that is odd. Therefore, a pure spin current exists 
only for $V_g=0$, while generically charge and spin currents both flow for 
non-zero gate voltages. Moreover, Fig.~\ref{fig6} exhibits a strong size 
dependence of the response to the gate voltage. In fact, the maxima for 
the curves for $I_s(V_g)$ are close to the highest (lowest) occupied 
(unoccupied) level that sits at the border of the shaded region, but still 
inside the gap. Since the single-particle gap decreases rapidly with 
increasing flake size $N$, for flakes with $N=54$ and $N=96$ sites, a 
small gate voltage results in small charge and spin currents, and a gate 
voltage of $\left|V_g/t\right|>0.2$ is required to induce appreciable 
charge and spin currents. By contrast, for the larger hexagonal-ZGNF 
sizes, namely $N=294 $ and $N=600$, a small gate voltage is sufficient to 
induce significant currents. In addition, for these larger flakes, the 
temperature $T=300\,{\rm K}$ leads to an appreciable current in the gaps 
owing to thermal fluctuations. As we have already seen in 
Fig.~\ref{fig4}(a--d), the maximum current increases with increasing $N$ 
(note the different scales of the vertical axes in Fig.~\ref{fig6}).


\section{Conclusion}
\label{secIV}

We have investigated the spin-caloritronic properties of hexagonal ZGNFs. 
First, we have treated the magnetic properties both by simple static MFT 
and the more accurate DMFT, providing us with the Green's functions and 
thus the transmission coefficients of the ZGNFs. The spin-dependent 
transport was then evaluated by the Landauer formalism in the framework of 
the non-equilibrium Green's function method. One of our main findings is 
that a pure spin current can be driven by a temperature gradient in the 
charge-neutral system if the Coulomb interaction is sufficiently large to 
render the zigzag edges magnetic, but still sufficiently small such that 
the bulk remains non-magnetic, and leads are attached to two edges with 
the same magnetization in a configuration that resembles the \textit{meta} 
binding motif in a benzene molecule. We found that the window of Coulomb 
interactions $U$ where only the edges are magnetic and thus a spin current 
flows increases significantly as the size of the flakes increases from 
$N=96$ to $N=600$, thus improving chances that the value of the Coulomb 
interaction in graphene is indeed in the relevant regime. At the same 
time, the absolute value of the spin current also increases significantly 
with increasing size of the flake and develops a maximum around room 
temperature for $N=600$. These observations promise room-temperature 
applications of ZGNFs with $N\approx 600$ carbon atoms.

The value of the spin current can be enhanced by an applied gate voltage 
$V_g$ with a significant effect in particular for the smaller flakes. We 
note that when the gate voltage exceeds the single-particle gap, electrons 
are doped into the system such that the magnetic state of the edges might 
change. Furthermore, doping reduces the gap values such that the effect of 
thermal fluctuations on the magnetic state may no longer be negligible. We 
suggest investigation of such effects as an interesting topic for further 
studies.

It should be mentioned that while dynamic properties are accurately 
captured by MFT, the static magnetism of the edge is actually an artifact 
of the mean-field approximation to the Hubbard model \cite{Feldner2011}. 
Further ingredients, such as a combination of spin-orbit coupling, 
external fields, or flakes with different edge lengths, will therefore be 
needed to stabilize magnetic edges in actual device applications. 
Furthermore, we have assumed transport to be ballistic and ignored all 
spin-relaxation and spin-flip processes. This is justified for a device 
which is smaller than the spin relaxation length. We also assume that all 
dissipation processes occur in the leads which remains reasonable for even the 
biggest flakes that we consider in this work. Thus, in order to use
devices in a spin-caloritronics application, a length
$L<l_{sp}\sim 1\mu\text{m}$ is needed, which is accessible in current 
experiments~\cite{Guimaraes2014}. Nevertheless, the thermoelectric effects 
in graphene-based devices have already been observed with considerable 
precision~\cite{Wei2009,Zuev2009}. On the other hand, a recent experiment 
used the local magnetoresistance technique to detect the thermoelectric spin 
voltage in graphene~\cite{Sierra2018}, which shows that in principle it is 
feasible to measure spin-dependent thermoelectric properties.

\begin{acknowledgments}

This work was supported by the Paris//Seine excellence initiative, the 911 
project of the Ministry of Education and Training of Vietnam, and the ANR 
project J2D (ANR-15-CE24-0017). R.P.\ is supported by JSPS KAKENHI Grants 
No.~18K03511 and No.~18H04316. The DMFT simulations were performed on the 
``Hokusai'' supercomputer in RIKEN and the supercomputer of the Institute 
for Solid State Physics (ISSP) in Japan. R.P.\ thanks the Universit\'e de 
Cergy-Pontoise and their Institute for Advanced Studies for hospitality 
during a research visit.

\end{acknowledgments}

\bibliography{graphenenanoflake}

\begin{thebibliography}{84}%
\makeatletter
\providecommand \@ifxundefined [1]{%
 \@ifx{#1\undefined}
}%
\providecommand \@ifnum [1]{%
 \ifnum #1\expandafter \@firstoftwo
 \else \expandafter \@secondoftwo
 \fi
}%
\providecommand \@ifx [1]{%
 \ifx #1\expandafter \@firstoftwo
 \else \expandafter \@secondoftwo
 \fi
}%
\providecommand \natexlab [1]{#1}%
\providecommand \enquote  [1]{``#1''}%
\providecommand \bibnamefont  [1]{#1}%
\providecommand \bibfnamefont [1]{#1}%
\providecommand \citenamefont [1]{#1}%
\providecommand \href@noop [0]{\@secondoftwo}%
\providecommand \href [0]{\begingroup \@sanitize@url \@href}%
\providecommand \@href[1]{\@@startlink{#1}\@@href}%
\providecommand \@@href[1]{\endgroup#1\@@endlink}%
\providecommand \@sanitize@url [0]{\catcode `\\12\catcode `\$12\catcode
  `\&12\catcode `\#12\catcode `\^12\catcode `\_12\catcode `\%12\relax}%
\providecommand \@@startlink[1]{}%
\providecommand \@@endlink[0]{}%
\providecommand \url  [0]{\begingroup\@sanitize@url \@url }%
\providecommand \@url [1]{\endgroup\@href {#1}{\urlprefix }}%
\providecommand \urlprefix  [0]{URL }%
\providecommand \Eprint [0]{\href }%
\providecommand \doibase [0]{http://dx.doi.org/}%
\providecommand \selectlanguage [0]{\@gobble}%
\providecommand \bibinfo  [0]{\@secondoftwo}%
\providecommand \bibfield  [0]{\@secondoftwo}%
\providecommand \translation [1]{[#1]}%
\providecommand \BibitemOpen [0]{}%
\providecommand \bibitemStop [0]{}%
\providecommand \bibitemNoStop [0]{.\EOS\space}%
\providecommand \EOS [0]{\spacefactor3000\relax}%
\providecommand \BibitemShut  [1]{\csname bibitem#1\endcsname}%
\let\auto@bib@innerbib\@empty
\bibitem [{\citenamefont {Kirihara}\ \emph {et~al.}(2012)\citenamefont
  {Kirihara}, \citenamefont {Uchida}, \citenamefont {Kajiwara}, \citenamefont
  {Ishida}, \citenamefont {Nakamura}, \citenamefont {Manako}, \citenamefont
  {Saitoh},\ and\ \citenamefont {Yorozu}}]{Kirihara2012}%
  \BibitemOpen
  \bibfield  {author} {\bibinfo {author} {\bibfnamefont {A.}~\bibnamefont
  {Kirihara}}, \bibinfo {author} {\bibfnamefont {K.}~\bibnamefont {Uchida}},
  \bibinfo {author} {\bibfnamefont {Y.}~\bibnamefont {Kajiwara}}, \bibinfo
  {author} {\bibfnamefont {M.}~\bibnamefont {Ishida}}, \bibinfo {author}
  {\bibfnamefont {Y.}~\bibnamefont {Nakamura}}, \bibinfo {author}
  {\bibfnamefont {T.}~\bibnamefont {Manako}}, \bibinfo {author} {\bibfnamefont
  {E.}~\bibnamefont {Saitoh}}, \ and\ \bibinfo {author} {\bibfnamefont
  {S.}~\bibnamefont {Yorozu}},\ }\bibfield  {title} {\enquote {\bibinfo {title}
  {Spin-current-driven thermoelectric coating},}\ }\href {\doibase
  10.1038/nmat3360} {\bibfield  {journal} {\bibinfo  {journal} {Nat. Mater.}\
  }\textbf {\bibinfo {volume} {11}},\ \bibinfo {pages} {686} (\bibinfo {year}
  {2012})}\BibitemShut {NoStop}%
\bibitem [{\citenamefont {Kim}\ \emph {et~al.}(2014)\citenamefont {Kim},
  \citenamefont {Jeong}, \citenamefont {Kim}, \citenamefont {Lee},\ and\
  \citenamefont {Reddy}}]{Kim2014}%
  \BibitemOpen
  \bibfield  {author} {\bibinfo {author} {\bibfnamefont {Y.}~\bibnamefont
  {Kim}}, \bibinfo {author} {\bibfnamefont {W.}~\bibnamefont {Jeong}}, \bibinfo
  {author} {\bibfnamefont {K.}~\bibnamefont {Kim}}, \bibinfo {author}
  {\bibfnamefont {W.}~\bibnamefont {Lee}}, \ and\ \bibinfo {author}
  {\bibfnamefont {P.}~\bibnamefont {Reddy}},\ }\bibfield  {title} {\enquote
  {\bibinfo {title} {Electrostatic control of thermoelectricity in molecular
  junctions},}\ }\href {\doibase 10.1038/nnano.2014.209} {\bibfield  {journal}
  {\bibinfo  {journal} {Nat. Nanotechnol.}\ }\textbf {\bibinfo {volume} {9}},\
  \bibinfo {pages} {881} (\bibinfo {year} {2014})}\BibitemShut {NoStop}%
\bibitem [{\citenamefont {Aviram}\ and\ \citenamefont
  {Ratner}(1974)}]{Aviram1974}%
  \BibitemOpen
  \bibfield  {author} {\bibinfo {author} {\bibfnamefont {A.}~\bibnamefont
  {Aviram}}\ and\ \bibinfo {author} {\bibfnamefont {M.~A.}\ \bibnamefont
  {Ratner}},\ }\bibfield  {title} {\enquote {\bibinfo {title} {Molecular
  rectifiers},}\ }\href {\doibase 10.1016/0009-2614(74)85031-1} {\bibfield
  {journal} {\bibinfo  {journal} {Chem. Phys. Lett.}\ }\textbf {\bibinfo
  {volume} {29}},\ \bibinfo {pages} {277} (\bibinfo {year} {1974})}\BibitemShut
  {NoStop}%
\bibitem [{\citenamefont {Pop}\ \emph {et~al.}(2006)\citenamefont {Pop},
  \citenamefont {Sinha},\ and\ \citenamefont {Goodson}}]{Pop2006}%
  \BibitemOpen
  \bibfield  {author} {\bibinfo {author} {\bibfnamefont {E.}~\bibnamefont
  {Pop}}, \bibinfo {author} {\bibfnamefont {S.}~\bibnamefont {Sinha}}, \ and\
  \bibinfo {author} {\bibfnamefont {K.~E.}\ \bibnamefont {Goodson}},\
  }\bibfield  {title} {\enquote {\bibinfo {title} {Heat generation and
  transport in nanometer-scale transistors},}\ }\href {\doibase
  10.1109/JPROC.2006.879794} {\bibfield  {journal} {\bibinfo  {journal} {Proc.
  IEEE}\ }\textbf {\bibinfo {volume} {94}},\ \bibinfo {pages} {1587} (\bibinfo
  {year} {2006})}\BibitemShut {NoStop}%
\bibitem [{\citenamefont {Wolf}\ \emph {et~al.}(2001)\citenamefont {Wolf},
  \citenamefont {Awschalom}, \citenamefont {Buhrman}, \citenamefont {Daughton},
  \citenamefont {von Moln{\'a}r}, \citenamefont {Roukes}, \citenamefont
  {Chtchelkanova},\ and\ \citenamefont {Treger}}]{Wolf2001}%
  \BibitemOpen
  \bibfield  {author} {\bibinfo {author} {\bibfnamefont {S.~A.}\ \bibnamefont
  {Wolf}}, \bibinfo {author} {\bibfnamefont {D.~D.}\ \bibnamefont {Awschalom}},
  \bibinfo {author} {\bibfnamefont {R.~A.}\ \bibnamefont {Buhrman}}, \bibinfo
  {author} {\bibfnamefont {J.~M.}\ \bibnamefont {Daughton}}, \bibinfo {author}
  {\bibfnamefont {S.}~\bibnamefont {von Moln{\'a}r}}, \bibinfo {author}
  {\bibfnamefont {M.~L.}\ \bibnamefont {Roukes}}, \bibinfo {author}
  {\bibfnamefont {A.~Y.}\ \bibnamefont {Chtchelkanova}}, \ and\ \bibinfo
  {author} {\bibfnamefont {D.~M.}\ \bibnamefont {Treger}},\ }\bibfield  {title}
  {\enquote {\bibinfo {title} {Spintronics: {A} spin-based electronics vision
  for the future},}\ }\href {\doibase 10.1126/science.1065389} {\bibfield
  {journal} {\bibinfo  {journal} {Science}\ }\textbf {\bibinfo {volume}
  {294}},\ \bibinfo {pages} {1488} (\bibinfo {year} {2001})}\BibitemShut
  {NoStop}%
\bibitem [{\citenamefont {Fert}(2008)}]{Fert2008}%
  \BibitemOpen
  \bibfield  {author} {\bibinfo {author} {\bibfnamefont {A.}~\bibnamefont
  {Fert}},\ }\bibfield  {title} {\enquote {\bibinfo {title} {Nobel lecture:
  Origin, development, and future of spintronics},}\ }\href {\doibase
  10.1103/RevModPhys.80.1517} {\bibfield  {journal} {\bibinfo  {journal} {Rev.
  Mod. Phys.}\ }\textbf {\bibinfo {volume} {80}},\ \bibinfo {pages} {1517}
  (\bibinfo {year} {2008})}\BibitemShut {NoStop}%
\bibitem [{\citenamefont {Uchida}\ \emph {et~al.}(2008)\citenamefont {Uchida},
  \citenamefont {Takahashi}, \citenamefont {Harii}, \citenamefont {Ieda},
  \citenamefont {Koshibae}, \citenamefont {Ando}, \citenamefont {Maekawa},\
  and\ \citenamefont {Saitoh}}]{Uchida2008}%
  \BibitemOpen
  \bibfield  {author} {\bibinfo {author} {\bibfnamefont {K.}~\bibnamefont
  {Uchida}}, \bibinfo {author} {\bibfnamefont {S.}~\bibnamefont {Takahashi}},
  \bibinfo {author} {\bibfnamefont {K.}~\bibnamefont {Harii}}, \bibinfo
  {author} {\bibfnamefont {J.}~\bibnamefont {Ieda}}, \bibinfo {author}
  {\bibfnamefont {W.}~\bibnamefont {Koshibae}}, \bibinfo {author}
  {\bibfnamefont {K.}~\bibnamefont {Ando}}, \bibinfo {author} {\bibfnamefont
  {S.}~\bibnamefont {Maekawa}}, \ and\ \bibinfo {author} {\bibfnamefont
  {E.}~\bibnamefont {Saitoh}},\ }\bibfield  {title} {\enquote {\bibinfo {title}
  {Observation of the spin {S}eebeck effect},}\ }\href {\doibase
  10.1038/nature07321} {\bibfield  {journal} {\bibinfo  {journal} {Nature}\
  }\textbf {\bibinfo {volume} {445}},\ \bibinfo {pages} {778} (\bibinfo {year}
  {2008})}\BibitemShut {NoStop}%
\bibitem [{\citenamefont {Adachi}\ \emph {et~al.}(2013)\citenamefont {Adachi},
  \citenamefont {Uchida}, \citenamefont {Saitoh},\ and\ \citenamefont
  {Maekawa}}]{Adachi2013}%
  \BibitemOpen
  \bibfield  {author} {\bibinfo {author} {\bibfnamefont {H.}~\bibnamefont
  {Adachi}}, \bibinfo {author} {\bibfnamefont {K.}~\bibnamefont {Uchida}},
  \bibinfo {author} {\bibfnamefont {E.}~\bibnamefont {Saitoh}}, \ and\ \bibinfo
  {author} {\bibfnamefont {S.}~\bibnamefont {Maekawa}},\ }\bibfield  {title}
  {\enquote {\bibinfo {title} {Theory of the spin {S}eebeck effect},}\ }\href
  {\doibase 10.1088/0034-4885/76/3/036501} {\bibfield  {journal} {\bibinfo
  {journal} {Rep. Prog. Phys.}\ }\textbf {\bibinfo {volume} {76}},\ \bibinfo
  {pages} {036501} (\bibinfo {year} {2013})}\BibitemShut {NoStop}%
\bibitem [{\citenamefont {Uchida}\ \emph {et~al.}(2010)\citenamefont {Uchida},
  \citenamefont {Xiao}, \citenamefont {Adachi}, \citenamefont {Ohe},
  \citenamefont {Takahashi}, \citenamefont {Ieda}, \citenamefont {Ota},
  \citenamefont {Kajiwara}, \citenamefont {Umezawa}, \citenamefont {Kawai},
  \citenamefont {Bauer}, \citenamefont {Maekawa},\ and\ \citenamefont
  {Saitoh}}]{Uchida2010}%
  \BibitemOpen
  \bibfield  {author} {\bibinfo {author} {\bibfnamefont {K.}~\bibnamefont
  {Uchida}}, \bibinfo {author} {\bibfnamefont {J.}~\bibnamefont {Xiao}},
  \bibinfo {author} {\bibfnamefont {H.}~\bibnamefont {Adachi}}, \bibinfo
  {author} {\bibfnamefont {J.}~\bibnamefont {Ohe}}, \bibinfo {author}
  {\bibfnamefont {S.}~\bibnamefont {Takahashi}}, \bibinfo {author}
  {\bibfnamefont {J.}~\bibnamefont {Ieda}}, \bibinfo {author} {\bibfnamefont
  {T.}~\bibnamefont {Ota}}, \bibinfo {author} {\bibfnamefont {Y.}~\bibnamefont
  {Kajiwara}}, \bibinfo {author} {\bibfnamefont {H.}~\bibnamefont {Umezawa}},
  \bibinfo {author} {\bibfnamefont {H.}~\bibnamefont {Kawai}}, \bibinfo
  {author} {\bibfnamefont {G.~E.~W.}\ \bibnamefont {Bauer}}, \bibinfo {author}
  {\bibfnamefont {S.}~\bibnamefont {Maekawa}}, \ and\ \bibinfo {author}
  {\bibfnamefont {E.}~\bibnamefont {Saitoh}},\ }\bibfield  {title} {\enquote
  {\bibinfo {title} {Spin {S}eebeck insulator},}\ }\href {\doibase
  10.1038/nmat2856} {\bibfield  {journal} {\bibinfo  {journal} {Nat. Mater.}\
  }\textbf {\bibinfo {volume} {9}},\ \bibinfo {pages} {894} (\bibinfo {year}
  {2010})}\BibitemShut {NoStop}%
\bibitem [{\citenamefont {Jaworski}\ \emph {et~al.}(2010)\citenamefont
  {Jaworski}, \citenamefont {Yang}, \citenamefont {Mack}, \citenamefont
  {Awschalom}, \citenamefont {Heremans},\ and\ \citenamefont
  {Myers}}]{Jaworski2010}%
  \BibitemOpen
  \bibfield  {author} {\bibinfo {author} {\bibfnamefont {C.~M.}\ \bibnamefont
  {Jaworski}}, \bibinfo {author} {\bibfnamefont {J.}~\bibnamefont {Yang}},
  \bibinfo {author} {\bibfnamefont {S.}~\bibnamefont {Mack}}, \bibinfo {author}
  {\bibfnamefont {D.~D.}\ \bibnamefont {Awschalom}}, \bibinfo {author}
  {\bibfnamefont {J.~P.}\ \bibnamefont {Heremans}}, \ and\ \bibinfo {author}
  {\bibfnamefont {R.~C.}\ \bibnamefont {Myers}},\ }\bibfield  {title} {\enquote
  {\bibinfo {title} {Observation of the spin-{S}eebeck effect in a
  ferromagnetic semiconductor},}\ }\href {\doibase 10.1038/nmat2860} {\bibfield
   {journal} {\bibinfo  {journal} {Nat. Mater.}\ }\textbf {\bibinfo {volume}
  {9}},\ \bibinfo {pages} {898} (\bibinfo {year} {2010})}\BibitemShut {NoStop}%
\bibitem [{\citenamefont {Bauer}\ \emph {et~al.}(2012)\citenamefont {Bauer},
  \citenamefont {Saitoh},\ and\ \citenamefont {van Wees}}]{Bauer2012}%
  \BibitemOpen
  \bibfield  {author} {\bibinfo {author} {\bibfnamefont {G.~E.~W.}\
  \bibnamefont {Bauer}}, \bibinfo {author} {\bibfnamefont {E.}~\bibnamefont
  {Saitoh}}, \ and\ \bibinfo {author} {\bibfnamefont {B.~J.}\ \bibnamefont {van
  Wees}},\ }\bibfield  {title} {\enquote {\bibinfo {title} {Spin
  caloritronics},}\ }\href {\doibase 10.1038/nmat3301} {\bibfield  {journal}
  {\bibinfo  {journal} {Nat. Mater.}\ }\textbf {\bibinfo {volume} {11}},\
  \bibinfo {pages} {391} (\bibinfo {year} {2012})}\BibitemShut {NoStop}%
\bibitem [{\citenamefont {Castro~Neto}\ \emph {et~al.}(2009)\citenamefont
  {Castro~Neto}, \citenamefont {Guinea}, \citenamefont {Peres}, \citenamefont
  {Novoselov},\ and\ \citenamefont {Geim}}]{CastroNeto2009}%
  \BibitemOpen
  \bibfield  {author} {\bibinfo {author} {\bibfnamefont {A.~H.}\ \bibnamefont
  {Castro~Neto}}, \bibinfo {author} {\bibfnamefont {F.}~\bibnamefont {Guinea}},
  \bibinfo {author} {\bibfnamefont {N.~M.~R.}\ \bibnamefont {Peres}}, \bibinfo
  {author} {\bibfnamefont {K.~S.}\ \bibnamefont {Novoselov}}, \ and\ \bibinfo
  {author} {\bibfnamefont {A.~K.}\ \bibnamefont {Geim}},\ }\bibfield  {title}
  {\enquote {\bibinfo {title} {The electronic properties of graphene},}\ }\href
  {\doibase 10.1103/RevModPhys.81.109} {\bibfield  {journal} {\bibinfo
  {journal} {Rev. Mod. Phys.}\ }\textbf {\bibinfo {volume} {81}},\ \bibinfo
  {pages} {109} (\bibinfo {year} {2009})}\BibitemShut {NoStop}%
\bibitem [{\citenamefont {Das~Sarma}\ \emph {et~al.}(2011)\citenamefont
  {Das~Sarma}, \citenamefont {Adam}, \citenamefont {Hwang},\ and\ \citenamefont
  {Rossi}}]{Sarma2011}%
  \BibitemOpen
  \bibfield  {author} {\bibinfo {author} {\bibfnamefont {S.}~\bibnamefont
  {Das~Sarma}}, \bibinfo {author} {\bibfnamefont {S.}~\bibnamefont {Adam}},
  \bibinfo {author} {\bibfnamefont {E.~H.}\ \bibnamefont {Hwang}}, \ and\
  \bibinfo {author} {\bibfnamefont {E.}~\bibnamefont {Rossi}},\ }\bibfield
  {title} {\enquote {\bibinfo {title} {Electronic transport in two-dimensional
  graphene},}\ }\href {\doibase 10.1103/RevModPhys.83.407} {\bibfield
  {journal} {\bibinfo  {journal} {Rev. Mod. Phys.}\ }\textbf {\bibinfo {volume}
  {83}},\ \bibinfo {pages} {407} (\bibinfo {year} {2011})}\BibitemShut
  {NoStop}%
\bibitem [{\citenamefont {Novoselov}\ \emph {et~al.}(2012)\citenamefont
  {Novoselov}, \citenamefont {Fal'ko}, \citenamefont {Colombo}, \citenamefont
  {Gellert}, \citenamefont {Schwab},\ and\ \citenamefont
  {Kim}}]{Novoselov2012}%
  \BibitemOpen
  \bibfield  {author} {\bibinfo {author} {\bibfnamefont {K.~S.}\ \bibnamefont
  {Novoselov}}, \bibinfo {author} {\bibfnamefont {V.~I.}\ \bibnamefont
  {Fal'ko}}, \bibinfo {author} {\bibfnamefont {L.}~\bibnamefont {Colombo}},
  \bibinfo {author} {\bibfnamefont {P.~R.}\ \bibnamefont {Gellert}}, \bibinfo
  {author} {\bibfnamefont {M.~G.}\ \bibnamefont {Schwab}}, \ and\ \bibinfo
  {author} {\bibfnamefont {K.}~\bibnamefont {Kim}},\ }\bibfield  {title}
  {\enquote {\bibinfo {title} {A roadmap for graphene},}\ }\href {\doibase
  10.1038/nature11458} {\bibfield  {journal} {\bibinfo  {journal} {Nature}\
  }\textbf {\bibinfo {volume} {490}},\ \bibinfo {pages} {192} (\bibinfo {year}
  {2012})}\BibitemShut {NoStop}%
\bibitem [{\citenamefont {Novoselov}\ \emph {et~al.}(2004)\citenamefont
  {Novoselov}, \citenamefont {Geim}, \citenamefont {Morozov}, \citenamefont
  {Jiang}, \citenamefont {Zhang}, \citenamefont {Dubonos}, \citenamefont
  {Grigorieva},\ and\ \citenamefont {Firsov}}]{Novoselov04}%
  \BibitemOpen
  \bibfield  {author} {\bibinfo {author} {\bibfnamefont {K.~S.}\ \bibnamefont
  {Novoselov}}, \bibinfo {author} {\bibfnamefont {A.~K.}\ \bibnamefont {Geim}},
  \bibinfo {author} {\bibfnamefont {S.~V.}\ \bibnamefont {Morozov}}, \bibinfo
  {author} {\bibfnamefont {D.}~\bibnamefont {Jiang}}, \bibinfo {author}
  {\bibfnamefont {Y.}~\bibnamefont {Zhang}}, \bibinfo {author} {\bibfnamefont
  {S.~V.}\ \bibnamefont {Dubonos}}, \bibinfo {author} {\bibfnamefont {I.~V.}\
  \bibnamefont {Grigorieva}}, \ and\ \bibinfo {author} {\bibfnamefont {A.~A.}\
  \bibnamefont {Firsov}},\ }\bibfield  {title} {\enquote {\bibinfo {title}
  {Electric field effect in atomically thin carbon films},}\ }\href {\doibase
  10.1126/science.1102896} {\bibfield  {journal} {\bibinfo  {journal}
  {Science}\ }\textbf {\bibinfo {volume} {306}},\ \bibinfo {pages} {666}
  (\bibinfo {year} {2004})}\BibitemShut {NoStop}%
\bibitem [{\citenamefont {Fujita}\ \emph {et~al.}(1996)\citenamefont {Fujita},
  \citenamefont {Wakabayashi}, \citenamefont {Nakada},\ and\ \citenamefont
  {Kusakabe}}]{Fujita96}%
  \BibitemOpen
  \bibfield  {author} {\bibinfo {author} {\bibfnamefont {M.}~\bibnamefont
  {Fujita}}, \bibinfo {author} {\bibfnamefont {K.}~\bibnamefont {Wakabayashi}},
  \bibinfo {author} {\bibfnamefont {K.}~\bibnamefont {Nakada}}, \ and\ \bibinfo
  {author} {\bibfnamefont {K.}~\bibnamefont {Kusakabe}},\ }\bibfield  {title}
  {\enquote {\bibinfo {title} {Peculiar localized state at zigzag graphite
  edge},}\ }\href {\doibase 10.1143/JPSJ.65.1920} {\bibfield  {journal}
  {\bibinfo  {journal} {J. Phys. Soc. Jpn.}\ }\textbf {\bibinfo {volume}
  {65}},\ \bibinfo {pages} {1920} (\bibinfo {year} {1996})}\BibitemShut
  {NoStop}%
\bibitem [{\citenamefont {Wakabayashi}\ \emph {et~al.}(1998)\citenamefont
  {Wakabayashi}, \citenamefont {Sigrist},\ and\ \citenamefont
  {Fujita}}]{Wakabayashi98}%
  \BibitemOpen
  \bibfield  {author} {\bibinfo {author} {\bibfnamefont {K.}~\bibnamefont
  {Wakabayashi}}, \bibinfo {author} {\bibfnamefont {M.}~\bibnamefont
  {Sigrist}}, \ and\ \bibinfo {author} {\bibfnamefont {M.}~\bibnamefont
  {Fujita}},\ }\bibfield  {title} {\enquote {\bibinfo {title} {Spin wave mode
  of edge-localized magnetic states in nanographite zigzag ribbons},}\ }\href
  {\doibase 10.1143/JPSJ.67.2089} {\bibfield  {journal} {\bibinfo  {journal}
  {J. Phys. Soc. Jpn.}\ }\textbf {\bibinfo {volume} {67}},\ \bibinfo {pages}
  {2089} (\bibinfo {year} {1998})}\BibitemShut {NoStop}%
\bibitem [{\citenamefont {Wakabayashi}\ \emph {et~al.}(1999)\citenamefont
  {Wakabayashi}, \citenamefont {Fujita}, \citenamefont {Ajiki},\ and\
  \citenamefont {Sigrist}}]{Wakabayashi99}%
  \BibitemOpen
  \bibfield  {author} {\bibinfo {author} {\bibfnamefont {K.}~\bibnamefont
  {Wakabayashi}}, \bibinfo {author} {\bibfnamefont {M.}~\bibnamefont {Fujita}},
  \bibinfo {author} {\bibfnamefont {H.}~\bibnamefont {Ajiki}}, \ and\ \bibinfo
  {author} {\bibfnamefont {M.}~\bibnamefont {Sigrist}},\ }\bibfield  {title}
  {\enquote {\bibinfo {title} {Electronic and magnetic properties of
  nanographite ribbons},}\ }\href {\doibase 10.1103/PhysRevB.59.8271}
  {\bibfield  {journal} {\bibinfo  {journal} {Phys. Rev. B}\ }\textbf {\bibinfo
  {volume} {59}},\ \bibinfo {pages} {8271} (\bibinfo {year}
  {1999})}\BibitemShut {NoStop}%
\bibitem [{\citenamefont {Nakada}\ \emph {et~al.}(1996)\citenamefont {Nakada},
  \citenamefont {Fujita}, \citenamefont {Dresselhaus},\ and\ \citenamefont
  {Dresselhaus}}]{Nakada96}%
  \BibitemOpen
  \bibfield  {author} {\bibinfo {author} {\bibfnamefont {K.}~\bibnamefont
  {Nakada}}, \bibinfo {author} {\bibfnamefont {M.}~\bibnamefont {Fujita}},
  \bibinfo {author} {\bibfnamefont {G.}~\bibnamefont {Dresselhaus}}, \ and\
  \bibinfo {author} {\bibfnamefont {M.~S.}\ \bibnamefont {Dresselhaus}},\
  }\bibfield  {title} {\enquote {\bibinfo {title} {Edge state in graphene
  ribbons: Nanometer size effect and edge shape dependence},}\ }\href {\doibase
  10.1103/PhysRevB.54.17954} {\bibfield  {journal} {\bibinfo  {journal} {Phys.
  Rev. B}\ }\textbf {\bibinfo {volume} {54}},\ \bibinfo {pages} {17954}
  (\bibinfo {year} {1996})}\BibitemShut {NoStop}%
\bibitem [{\citenamefont {Brey}\ and\ \citenamefont {Fertig}(2006)}]{Brey06}%
  \BibitemOpen
  \bibfield  {author} {\bibinfo {author} {\bibfnamefont {L.}~\bibnamefont
  {Brey}}\ and\ \bibinfo {author} {\bibfnamefont {H.~A.}\ \bibnamefont
  {Fertig}},\ }\bibfield  {title} {\enquote {\bibinfo {title} {Electronic
  states of graphene nanoribbons studied with the {D}irac equation},}\ }\href
  {\doibase 10.1103/PhysRevB.73.235411} {\bibfield  {journal} {\bibinfo
  {journal} {Phys. Rev. B}\ }\textbf {\bibinfo {volume} {73}},\ \bibinfo
  {pages} {235411} (\bibinfo {year} {2006})}\BibitemShut {NoStop}%
\bibitem [{\citenamefont {Zuev}\ \emph {et~al.}(2009)\citenamefont {Zuev},
  \citenamefont {Chang},\ and\ \citenamefont {Kim}}]{Zuev2009}%
  \BibitemOpen
  \bibfield  {author} {\bibinfo {author} {\bibfnamefont {Y.~M.}\ \bibnamefont
  {Zuev}}, \bibinfo {author} {\bibfnamefont {W.}~\bibnamefont {Chang}}, \ and\
  \bibinfo {author} {\bibfnamefont {P.}~\bibnamefont {Kim}},\ }\bibfield
  {title} {\enquote {\bibinfo {title} {Thermoelectric and magnetothermoelectric
  transport measurements of graphene},}\ }\href {\doibase
  10.1103/PhysRevLett.102.096807} {\bibfield  {journal} {\bibinfo  {journal}
  {Phys. Rev. Lett.}\ }\textbf {\bibinfo {volume} {102}},\ \bibinfo {pages}
  {096807} (\bibinfo {year} {2009})}\BibitemShut {NoStop}%
\bibitem [{\citenamefont {Wei}\ \emph {et~al.}(2009)\citenamefont {Wei},
  \citenamefont {Bao}, \citenamefont {Pu}, \citenamefont {Lau},\ and\
  \citenamefont {Shi}}]{Wei2009}%
  \BibitemOpen
  \bibfield  {author} {\bibinfo {author} {\bibfnamefont {P.}~\bibnamefont
  {Wei}}, \bibinfo {author} {\bibfnamefont {W.}~\bibnamefont {Bao}}, \bibinfo
  {author} {\bibfnamefont {Y.}~\bibnamefont {Pu}}, \bibinfo {author}
  {\bibfnamefont {C.~N.}\ \bibnamefont {Lau}}, \ and\ \bibinfo {author}
  {\bibfnamefont {J.}~\bibnamefont {Shi}},\ }\bibfield  {title} {\enquote
  {\bibinfo {title} {Anomalous thermoelectric transport of {D}irac particles in
  graphene},}\ }\href {\doibase 10.1103/PhysRevLett.102.166808} {\bibfield
  {journal} {\bibinfo  {journal} {Phys. Rev. Lett.}\ }\textbf {\bibinfo
  {volume} {102}},\ \bibinfo {pages} {166808} (\bibinfo {year}
  {2009})}\BibitemShut {NoStop}%
\bibitem [{\citenamefont {Checkelsky}\ and\ \citenamefont
  {Ong}(2009)}]{Checkelsky2009}%
  \BibitemOpen
  \bibfield  {author} {\bibinfo {author} {\bibfnamefont {J.~G.}\ \bibnamefont
  {Checkelsky}}\ and\ \bibinfo {author} {\bibfnamefont {N.~P.}\ \bibnamefont
  {Ong}},\ }\bibfield  {title} {\enquote {\bibinfo {title} {Thermopower and
  {N}ernst effect in graphene in a magnetic field},}\ }\href {\doibase
  10.1103/PhysRevB.80.081413} {\bibfield  {journal} {\bibinfo  {journal} {Phys.
  Rev. B}\ }\textbf {\bibinfo {volume} {80}},\ \bibinfo {pages} {081413(R)}
  (\bibinfo {year} {2009})}\BibitemShut {NoStop}%
\bibitem [{\citenamefont {Ghahari}\ \emph {et~al.}(2016)\citenamefont
  {Ghahari}, \citenamefont {Xie}, \citenamefont {Taniguchi}, \citenamefont
  {Watanabe}, \citenamefont {Foster},\ and\ \citenamefont {Kim}}]{Ghahari2016}%
  \BibitemOpen
  \bibfield  {author} {\bibinfo {author} {\bibfnamefont {F.}~\bibnamefont
  {Ghahari}}, \bibinfo {author} {\bibfnamefont {H.-Y.}\ \bibnamefont {Xie}},
  \bibinfo {author} {\bibfnamefont {T.}~\bibnamefont {Taniguchi}}, \bibinfo
  {author} {\bibfnamefont {K.}~\bibnamefont {Watanabe}}, \bibinfo {author}
  {\bibfnamefont {M.~S.}\ \bibnamefont {Foster}}, \ and\ \bibinfo {author}
  {\bibfnamefont {P.}~\bibnamefont {Kim}},\ }\bibfield  {title} {\enquote
  {\bibinfo {title} {Enhanced thermoelectric power in graphene: {V}iolation of
  the {M}ott relation by inelastic scattering},}\ }\href {\doibase
  10.1103/PhysRevLett.116.136802} {\bibfield  {journal} {\bibinfo  {journal}
  {Phys. Rev. Lett.}\ }\textbf {\bibinfo {volume} {116}},\ \bibinfo {pages}
  {136802} (\bibinfo {year} {2016})}\BibitemShut {NoStop}%
\bibitem [{\citenamefont {Vahedi}\ and\ \citenamefont
  {Barimani}(2016)}]{Vahedi2016}%
  \BibitemOpen
  \bibfield  {author} {\bibinfo {author} {\bibfnamefont {J.}~\bibnamefont
  {Vahedi}}\ and\ \bibinfo {author} {\bibfnamefont {F.}~\bibnamefont
  {Barimani}},\ }\bibfield  {title} {\enquote {\bibinfo {title} {Spin and
  charge thermopower effects in the ferromagnetic graphene junction},}\ }\href
  {\doibase 10.1063/1.4961093} {\bibfield  {journal} {\bibinfo  {journal} {J.
  Appl. Phys.}\ }\textbf {\bibinfo {volume} {120}},\ \bibinfo {pages} {084303}
  (\bibinfo {year} {2016})}\BibitemShut {NoStop}%
\bibitem [{\citenamefont {Saiz-Bret\'{\i}n}\ \emph {et~al.}(2015)\citenamefont
  {Saiz-Bret\'{\i}n}, \citenamefont {Malyshev}, \citenamefont {Orellana},\ and\
  \citenamefont {Dom\'{\i}nguez-Adame}}]{Bretin2015}%
  \BibitemOpen
  \bibfield  {author} {\bibinfo {author} {\bibfnamefont {M.}~\bibnamefont
  {Saiz-Bret\'{\i}n}}, \bibinfo {author} {\bibfnamefont {A.~V.}\ \bibnamefont
  {Malyshev}}, \bibinfo {author} {\bibfnamefont {P.~A.}\ \bibnamefont
  {Orellana}}, \ and\ \bibinfo {author} {\bibfnamefont {F.}~\bibnamefont
  {Dom\'{\i}nguez-Adame}},\ }\bibfield  {title} {\enquote {\bibinfo {title}
  {Enhancing thermoelectric properties of graphene quantum rings},}\ }\href
  {\doibase 10.1103/PhysRevB.91.085431} {\bibfield  {journal} {\bibinfo
  {journal} {Phys. Rev. B}\ }\textbf {\bibinfo {volume} {91}},\ \bibinfo
  {pages} {085431} (\bibinfo {year} {2015})}\BibitemShut {NoStop}%
\bibitem [{\citenamefont {Chico}\ \emph {et~al.}(2017)\citenamefont {Chico},
  \citenamefont {Orellana}, \citenamefont {Rosales},\ and\ \citenamefont
  {Pacheco}}]{Chico2017}%
  \BibitemOpen
  \bibfield  {author} {\bibinfo {author} {\bibfnamefont {L.}~\bibnamefont
  {Chico}}, \bibinfo {author} {\bibfnamefont {P.~A.}\ \bibnamefont {Orellana}},
  \bibinfo {author} {\bibfnamefont {L.}~\bibnamefont {Rosales}}, \ and\
  \bibinfo {author} {\bibfnamefont {M.}~\bibnamefont {Pacheco}},\ }\bibfield
  {title} {\enquote {\bibinfo {title} {Spin and charge caloritronics in bilayer
  graphene flakes with magnetic contacts},}\ }\href {\doibase
  10.1103/PhysRevApplied.8.054029} {\bibfield  {journal} {\bibinfo  {journal}
  {Phys. Rev. Applied}\ }\textbf {\bibinfo {volume} {8}},\ \bibinfo {pages}
  {054029} (\bibinfo {year} {2017})}\BibitemShut {NoStop}%
\bibitem [{\citenamefont {Zeng}\ \emph {et~al.}(2011)\citenamefont {Zeng},
  \citenamefont {Feng},\ and\ \citenamefont {Liang}}]{Zeng2011}%
  \BibitemOpen
  \bibfield  {author} {\bibinfo {author} {\bibfnamefont {M.}~\bibnamefont
  {Zeng}}, \bibinfo {author} {\bibfnamefont {Y.}~\bibnamefont {Feng}}, \ and\
  \bibinfo {author} {\bibfnamefont {G.}~\bibnamefont {Liang}},\ }\bibfield
  {title} {\enquote {\bibinfo {title} {Graphene-based spin caloritronics},}\
  }\href {\doibase 10.1021/nl2000049} {\bibfield  {journal} {\bibinfo
  {journal} {Nano Lett.}\ }\textbf {\bibinfo {volume} {11}},\ \bibinfo {pages}
  {1369} (\bibinfo {year} {2011})}\BibitemShut {NoStop}%
\bibitem [{\citenamefont {Zhao}\ \emph {et~al.}(2012)\citenamefont {Zhao},
  \citenamefont {Zhai},\ and\ \citenamefont {Jin}}]{Zhao2012}%
  \BibitemOpen
  \bibfield  {author} {\bibinfo {author} {\bibfnamefont {Z.}~\bibnamefont
  {Zhao}}, \bibinfo {author} {\bibfnamefont {X.}~\bibnamefont {Zhai}}, \ and\
  \bibinfo {author} {\bibfnamefont {G.}~\bibnamefont {Jin}},\ }\bibfield
  {title} {\enquote {\bibinfo {title} {Bipolar-unipolar transition in
  thermospin transport through a graphene-based transistor},}\ }\href {\doibase
  10.1063/1.4748110} {\bibfield  {journal} {\bibinfo  {journal} {Appl. Phys.
  Lett.}\ }\textbf {\bibinfo {volume} {101}},\ \bibinfo {pages} {083117}
  (\bibinfo {year} {2012})}\BibitemShut {NoStop}%
\bibitem [{\citenamefont {Ni}\ \emph {et~al.}(2013)\citenamefont {Ni},
  \citenamefont {Yao}, \citenamefont {Fu}, \citenamefont {Gao}, \citenamefont
  {Zhu},\ and\ \citenamefont {Wang}}]{Ni2013}%
  \BibitemOpen
  \bibfield  {author} {\bibinfo {author} {\bibfnamefont {Y.}~\bibnamefont
  {Ni}}, \bibinfo {author} {\bibfnamefont {K.}~\bibnamefont {Yao}}, \bibinfo
  {author} {\bibfnamefont {H..}\ \bibnamefont {Fu}}, \bibinfo {author}
  {\bibfnamefont {G.}~\bibnamefont {Gao}}, \bibinfo {author} {\bibfnamefont
  {S.}~\bibnamefont {Zhu}}, \ and\ \bibinfo {author} {\bibfnamefont
  {S.}~\bibnamefont {Wang}},\ }\bibfield  {title} {\enquote {\bibinfo {title}
  {Spin {S}eebeck effect and thermal colossal magnetoresistance in graphene
  nanoribbon heterojunction},}\ }\href {\doibase 10.1038/srep01380} {\bibfield
  {journal} {\bibinfo  {journal} {Sci. Rep.}\ }\textbf {\bibinfo {volume}
  {3}},\ \bibinfo {pages} {1380} (\bibinfo {year} {2013})}\BibitemShut
  {NoStop}%
\bibitem [{\citenamefont {Chen}\ \emph {et~al.}(2014)\citenamefont {Chen},
  \citenamefont {Liu}, \citenamefont {Gu}, \citenamefont {Duan},\ and\
  \citenamefont {Liu}}]{Chen2014}%
  \BibitemOpen
  \bibfield  {author} {\bibinfo {author} {\bibfnamefont {X.}~\bibnamefont
  {Chen}}, \bibinfo {author} {\bibfnamefont {Y.}~\bibnamefont {Liu}}, \bibinfo
  {author} {\bibfnamefont {B.-L.}\ \bibnamefont {Gu}}, \bibinfo {author}
  {\bibfnamefont {W.}~\bibnamefont {Duan}}, \ and\ \bibinfo {author}
  {\bibfnamefont {F.}~\bibnamefont {Liu}},\ }\bibfield  {title} {\enquote
  {\bibinfo {title} {Giant room-temperature spin caloritronics in
  spin-semiconducting graphene nanoribbons},}\ }\href {\doibase
  10.1103/PhysRevB.90.121403} {\bibfield  {journal} {\bibinfo  {journal} {Phys.
  Rev. B}\ }\textbf {\bibinfo {volume} {90}},\ \bibinfo {pages} {121403(R)}
  (\bibinfo {year} {2014})}\BibitemShut {NoStop}%
\bibitem [{\citenamefont {Liang}\ \emph {et~al.}(2012)\citenamefont {Liang},
  \citenamefont {Cruz-Silva}, \citenamefont {Gir\~ao},\ and\ \citenamefont
  {Meunier}}]{Liang2012}%
  \BibitemOpen
  \bibfield  {author} {\bibinfo {author} {\bibfnamefont {L.}~\bibnamefont
  {Liang}}, \bibinfo {author} {\bibfnamefont {E.}~\bibnamefont {Cruz-Silva}},
  \bibinfo {author} {\bibfnamefont {E.~C.}\ \bibnamefont {Gir\~ao}}, \ and\
  \bibinfo {author} {\bibfnamefont {V.}~\bibnamefont {Meunier}},\ }\bibfield
  {title} {\enquote {\bibinfo {title} {Enhanced thermoelectric figure of merit
  in assembled graphene nanoribbons},}\ }\href {\doibase
  10.1103/PhysRevB.86.115438} {\bibfield  {journal} {\bibinfo  {journal} {Phys.
  Rev. B}\ }\textbf {\bibinfo {volume} {86}},\ \bibinfo {pages} {115438}
  (\bibinfo {year} {2012})}\BibitemShut {NoStop}%
\bibitem [{\citenamefont {Shirdel-Havar}\ and\ \citenamefont
  {Farghadan}(2018)}]{Farghadan2018}%
  \BibitemOpen
  \bibfield  {author} {\bibinfo {author} {\bibfnamefont {M.}~\bibnamefont
  {Shirdel-Havar}}\ and\ \bibinfo {author} {\bibfnamefont {R.}~\bibnamefont
  {Farghadan}},\ }\bibfield  {title} {\enquote {\bibinfo {title} {Spin
  caloritronics in spin semiconducting armchair graphene nanoribbons},}\ }\href
  {\doibase 10.1103/PhysRevB.97.235421} {\bibfield  {journal} {\bibinfo
  {journal} {Phys. Rev. B}\ }\textbf {\bibinfo {volume} {97}},\ \bibinfo
  {pages} {235421} (\bibinfo {year} {2018})}\BibitemShut {NoStop}%
\bibitem [{\citenamefont {Zberecki}\ \emph {et~al.}(2013)\citenamefont
  {Zberecki}, \citenamefont {Wierzbicki}, \citenamefont
  {Barna\ifmmode~\acute{s}\else \'{s}\fi{}},\ and\ \citenamefont
  {Swirkowicz}}]{Zberecki2013}%
  \BibitemOpen
  \bibfield  {author} {\bibinfo {author} {\bibfnamefont {K.}~\bibnamefont
  {Zberecki}}, \bibinfo {author} {\bibfnamefont {M.}~\bibnamefont
  {Wierzbicki}}, \bibinfo {author} {\bibfnamefont {J.}~\bibnamefont
  {Barna\ifmmode~\acute{s}\else \'{s}\fi{}}}, \ and\ \bibinfo {author}
  {\bibfnamefont {R.}~\bibnamefont {Swirkowicz}},\ }\bibfield  {title}
  {\enquote {\bibinfo {title} {Thermoelectric effects in silicene
  nanoribbons},}\ }\href {\doibase 10.1103/PhysRevB.88.115404} {\bibfield
  {journal} {\bibinfo  {journal} {Phys. Rev. B}\ }\textbf {\bibinfo {volume}
  {88}},\ \bibinfo {pages} {115404} (\bibinfo {year} {2013})}\BibitemShut
  {NoStop}%
\bibitem [{\citenamefont {Zhai}\ \emph {et~al.}(2014)\citenamefont {Zhai},
  \citenamefont {Wang}, \citenamefont {Vasilopoulos}, \citenamefont {Liu},
  \citenamefont {Dong}, \citenamefont {Zhou}, \citenamefont {Jiang},\ and\
  \citenamefont {You}}]{Zhai2014}%
  \BibitemOpen
  \bibfield  {author} {\bibinfo {author} {\bibfnamefont {M.-X.}\ \bibnamefont
  {Zhai}}, \bibinfo {author} {\bibfnamefont {X.-F.}\ \bibnamefont {Wang}},
  \bibinfo {author} {\bibfnamefont {P.}~\bibnamefont {Vasilopoulos}}, \bibinfo
  {author} {\bibfnamefont {Y.-S.}\ \bibnamefont {Liu}}, \bibinfo {author}
  {\bibfnamefont {Y.-J.}\ \bibnamefont {Dong}}, \bibinfo {author}
  {\bibfnamefont {L.}~\bibnamefont {Zhou}}, \bibinfo {author} {\bibfnamefont
  {Y.-J.}\ \bibnamefont {Jiang}}, \ and\ \bibinfo {author} {\bibfnamefont
  {W.-L.}\ \bibnamefont {You}},\ }\bibfield  {title} {\enquote {\bibinfo
  {title} {Giant magnetoresistance and spin {S}eebeck coefficient in zigzag
  $\alpha$-graphyne nanoribbons},}\ }\href {\doibase 10.1039/C4NR02426E}
  {\bibfield  {journal} {\bibinfo  {journal} {Nanoscale}\ }\textbf {\bibinfo
  {volume} {6}},\ \bibinfo {pages} {11121} (\bibinfo {year}
  {2014})}\BibitemShut {NoStop}%
\bibitem [{\citenamefont {Li}\ \emph {et~al.}(2016)\citenamefont {Li},
  \citenamefont {Wang}, \citenamefont {Xu}, \citenamefont {Wei},\ and\
  \citenamefont {Wang}}]{Li2016}%
  \BibitemOpen
  \bibfield  {author} {\bibinfo {author} {\bibfnamefont {J.~W.}\ \bibnamefont
  {Li}}, \bibinfo {author} {\bibfnamefont {B.}~\bibnamefont {Wang}}, \bibinfo
  {author} {\bibfnamefont {F.~M.}\ \bibnamefont {Xu}}, \bibinfo {author}
  {\bibfnamefont {Y.~D.}\ \bibnamefont {Wei}}, \ and\ \bibinfo {author}
  {\bibfnamefont {J.}~\bibnamefont {Wang}},\ }\bibfield  {title} {\enquote
  {\bibinfo {title} {Spin-dependent {S}eebeck effects in graphene-based
  molecular junctions},}\ }\href {\doibase 10.1103/PhysRevB.93.195426}
  {\bibfield  {journal} {\bibinfo  {journal} {Phys. Rev. B}\ }\textbf {\bibinfo
  {volume} {93}},\ \bibinfo {pages} {195426} (\bibinfo {year}
  {2016})}\BibitemShut {NoStop}%
\bibitem [{\citenamefont {Yazyev}(2010)}]{Yazyev2010}%
  \BibitemOpen
  \bibfield  {author} {\bibinfo {author} {\bibfnamefont {O.~V.}\ \bibnamefont
  {Yazyev}},\ }\bibfield  {title} {\enquote {\bibinfo {title} {Emergence of
  magnetism in graphene materials and nanostructures},}\ }\href {\doibase
  10.1088/0034-4885/73/5/056501} {\bibfield  {journal} {\bibinfo  {journal}
  {Rep. Prog. Phys.}\ }\textbf {\bibinfo {volume} {73}},\ \bibinfo {pages}
  {056501} (\bibinfo {year} {2010})}\BibitemShut {NoStop}%
\bibitem [{\citenamefont {Dong}\ \emph {et~al.}(2019)\citenamefont {Dong},
  \citenamefont {Wu}, \citenamefont {Wang}, \citenamefont {Yang},\ and\
  \citenamefont {Liu}}]{Dong2019}%
  \BibitemOpen
  \bibfield  {author} {\bibinfo {author} {\bibfnamefont {Y.}~\bibnamefont
  {Dong}}, \bibinfo {author} {\bibfnamefont {Y.}~\bibnamefont {Wu}}, \bibinfo
  {author} {\bibfnamefont {X.}~\bibnamefont {Wang}}, \bibinfo {author}
  {\bibfnamefont {X.}~\bibnamefont {Yang}}, \ and\ \bibinfo {author}
  {\bibfnamefont {Y.}~\bibnamefont {Liu}},\ }\bibfield  {title} {\enquote
  {\bibinfo {title} {Nanoporous graphene nanoribbons: Robust
  spin-semiconducting property and perfect spin {S}eebeck effects},}\ }\href
  {\doibase 10.1021/acs.jpcc.9b08092} {\bibfield  {journal} {\bibinfo
  {journal} {J. Phys. Chem. C}\ }\textbf {\bibinfo {volume} {123}},\ \bibinfo
  {pages} {29126} (\bibinfo {year} {2019})}\BibitemShut {NoStop}%
\bibitem [{\citenamefont {Song}\ \emph {et~al.}(2020)\citenamefont {Song},
  \citenamefont {Jin}, \citenamefont {Liu}, \citenamefont {Yuan}, \citenamefont
  {Yang}, \citenamefont {Jiang},\ and\ \citenamefont {Zheng}}]{Song2020}%
  \BibitemOpen
  \bibfield  {author} {\bibinfo {author} {\bibfnamefont {L.}~\bibnamefont
  {Song}}, \bibinfo {author} {\bibfnamefont {S.}~\bibnamefont {Jin}}, \bibinfo
  {author} {\bibfnamefont {Y.}~\bibnamefont {Liu}}, \bibinfo {author}
  {\bibfnamefont {L.}~\bibnamefont {Yuan}}, \bibinfo {author} {\bibfnamefont
  {Z.}~\bibnamefont {Yang}}, \bibinfo {author} {\bibfnamefont {P.}~\bibnamefont
  {Jiang}}, \ and\ \bibinfo {author} {\bibfnamefont {X.}~\bibnamefont
  {Zheng}},\ }\bibfield  {title} {\enquote {\bibinfo {title} {Thermal gradient
  driven spin current in {BN} co-doped ferromagnetic zigzag graphene
  nanoribbons},}\ }\href {\doibase 10.1016/j.physe.2019.113684} {\bibfield
  {journal} {\bibinfo  {journal} {Physica E}\ }\textbf {\bibinfo {volume}
  {115}},\ \bibinfo {pages} {113684} (\bibinfo {year} {2020})}\BibitemShut
  {NoStop}%
\bibitem [{\citenamefont {Tan}\ \emph {et~al.}(2020)\citenamefont {Tan},
  \citenamefont {Yang}, \citenamefont {Yang},\ and\ \citenamefont
  {Liu}}]{Tan2020}%
  \BibitemOpen
  \bibfield  {author} {\bibinfo {author} {\bibfnamefont {F.~X.}\ \bibnamefont
  {Tan}}, \bibinfo {author} {\bibfnamefont {L.~Y.}\ \bibnamefont {Yang}},
  \bibinfo {author} {\bibfnamefont {X.~F.}\ \bibnamefont {Yang}}, \ and\
  \bibinfo {author} {\bibfnamefont {Y.~S.}\ \bibnamefont {Liu}},\ }\bibfield
  {title} {\enquote {\bibinfo {title} {Thermoelectric transport properties of
  magnetic carbon-based organic chains},}\ }\href {\doibase
  10.1016/j.chemphys.2019.110524} {\bibfield  {journal} {\bibinfo  {journal}
  {Chem. Phys.}\ }\textbf {\bibinfo {volume} {528}},\ \bibinfo {pages} {110524}
  (\bibinfo {year} {2020})}\BibitemShut {NoStop}%
\bibitem [{\citenamefont {Fern\'andez-Rossier}\ and\ \citenamefont
  {Palacios}(2007)}]{Fernandez2007}%
  \BibitemOpen
  \bibfield  {author} {\bibinfo {author} {\bibfnamefont {J.}~\bibnamefont
  {Fern\'andez-Rossier}}\ and\ \bibinfo {author} {\bibfnamefont {J.~J.}\
  \bibnamefont {Palacios}},\ }\bibfield  {title} {\enquote {\bibinfo {title}
  {Magnetism in graphene nanoislands},}\ }\href {\doibase
  10.1103/PhysRevLett.99.177204} {\bibfield  {journal} {\bibinfo  {journal}
  {Phys. Rev. Lett.}\ }\textbf {\bibinfo {volume} {99}},\ \bibinfo {pages}
  {177204} (\bibinfo {year} {2007})}\BibitemShut {NoStop}%
\bibitem [{\citenamefont {Bhowmick}\ and\ \citenamefont
  {Shenoy}(2008)}]{Bhowmick2008}%
  \BibitemOpen
  \bibfield  {author} {\bibinfo {author} {\bibfnamefont {S.}~\bibnamefont
  {Bhowmick}}\ and\ \bibinfo {author} {\bibfnamefont {V.~B.}\ \bibnamefont
  {Shenoy}},\ }\bibfield  {title} {\enquote {\bibinfo {title} {Edge state
  magnetism of single layer graphene nanostructures},}\ }\href {\doibase
  10.1063/1.2943678} {\bibfield  {journal} {\bibinfo  {journal} {J. Chem.
  Phys.}\ }\textbf {\bibinfo {volume} {128}},\ \bibinfo {pages} {244717}
  (\bibinfo {year} {2008})}\BibitemShut {NoStop}%
\bibitem [{\citenamefont {Viana-Gomes}\ \emph {et~al.}(2009)\citenamefont
  {Viana-Gomes}, \citenamefont {Pereira},\ and\ \citenamefont
  {Peres}}]{Viana2009}%
  \BibitemOpen
  \bibfield  {author} {\bibinfo {author} {\bibfnamefont {J.}~\bibnamefont
  {Viana-Gomes}}, \bibinfo {author} {\bibfnamefont {V.~M.}\ \bibnamefont
  {Pereira}}, \ and\ \bibinfo {author} {\bibfnamefont {N.~M.~R.}\ \bibnamefont
  {Peres}},\ }\bibfield  {title} {\enquote {\bibinfo {title} {Magnetism in
  strained graphene dots},}\ }\href {\doibase 10.1103/PhysRevB.80.245436}
  {\bibfield  {journal} {\bibinfo  {journal} {Phys. Rev. B}\ }\textbf {\bibinfo
  {volume} {80}},\ \bibinfo {pages} {245436} (\bibinfo {year}
  {2009})}\BibitemShut {NoStop}%
\bibitem [{\citenamefont {Feldner}\ \emph {et~al.}(2010)\citenamefont
  {Feldner}, \citenamefont {Meng}, \citenamefont {Honecker}, \citenamefont
  {Cabra}, \citenamefont {Wessel},\ and\ \citenamefont {Assaad}}]{Feldner2010}%
  \BibitemOpen
  \bibfield  {author} {\bibinfo {author} {\bibfnamefont {H.}~\bibnamefont
  {Feldner}}, \bibinfo {author} {\bibfnamefont {Z.~Y.}\ \bibnamefont {Meng}},
  \bibinfo {author} {\bibfnamefont {A.}~\bibnamefont {Honecker}}, \bibinfo
  {author} {\bibfnamefont {D.}~\bibnamefont {Cabra}}, \bibinfo {author}
  {\bibfnamefont {S.}~\bibnamefont {Wessel}}, \ and\ \bibinfo {author}
  {\bibfnamefont {F.~F.}\ \bibnamefont {Assaad}},\ }\bibfield  {title}
  {\enquote {\bibinfo {title} {Magnetism of finite graphene samples: Mean-field
  theory compared with exact diagonalization and quantum {M}onte {C}arlo
  simulations},}\ }\href {\doibase 10.1103/PhysRevB.81.115416} {\bibfield
  {journal} {\bibinfo  {journal} {Phys. Rev. B}\ }\textbf {\bibinfo {volume}
  {81}},\ \bibinfo {pages} {115416} (\bibinfo {year} {2010})}\BibitemShut
  {NoStop}%
\bibitem [{\citenamefont {Feldner}\ \emph {et~al.}(2020)\citenamefont
  {Feldner}, \citenamefont {Meng}, \citenamefont {Honecker}, \citenamefont
  {Cabra}, \citenamefont {Wessel},\ and\ \citenamefont {Assaad}}]{FeldnerE}%
  \BibitemOpen
  \bibfield  {author} {\bibinfo {author} {\bibfnamefont {H.}~\bibnamefont
  {Feldner}}, \bibinfo {author} {\bibfnamefont {Z.~Y.}\ \bibnamefont {Meng}},
  \bibinfo {author} {\bibfnamefont {A.}~\bibnamefont {Honecker}}, \bibinfo
  {author} {\bibfnamefont {D.}~\bibnamefont {Cabra}}, \bibinfo {author}
  {\bibfnamefont {S.}~\bibnamefont {Wessel}}, \ and\ \bibinfo {author}
  {\bibfnamefont {F.~F.}\ \bibnamefont {Assaad}},\ }\bibfield  {title}
  {{\bibinfo {title} {Erratum:}}\ }\href {\doibase
  10.1103/PhysRevB.101.049909} {\bibfield  {journal} {\bibinfo  {journal}
  {Phys. Rev. B}\ }\textbf {\bibinfo {volume} {101}},\ \bibinfo {pages}
  {049909(E)} (\bibinfo {year} {2020})}\BibitemShut {NoStop}%
\bibitem [{\citenamefont {Roy}\ \emph {et~al.}(2014)\citenamefont {Roy},
  \citenamefont {Assaad},\ and\ \citenamefont {Herbut}}]{Roy2014}%
  \BibitemOpen
  \bibfield  {author} {\bibinfo {author} {\bibfnamefont {B.}~\bibnamefont
  {Roy}}, \bibinfo {author} {\bibfnamefont {F.~F.}\ \bibnamefont {Assaad}}, \
  and\ \bibinfo {author} {\bibfnamefont {I.~F.}\ \bibnamefont {Herbut}},\
  }\bibfield  {title} {\enquote {\bibinfo {title} {Zero modes and global
  antiferromagnetism in strained graphene},}\ }\href {\doibase
  10.1103/PhysRevX.4.021042} {\bibfield  {journal} {\bibinfo  {journal} {Phys.
  Rev. X}\ }\textbf {\bibinfo {volume} {4}},\ \bibinfo {pages} {021042}
  (\bibinfo {year} {2014})}\BibitemShut {NoStop}%
\bibitem [{\citenamefont {Valli}\ \emph {et~al.}(2016)\citenamefont {Valli},
  \citenamefont {Amaricci}, \citenamefont {Toschi}, \citenamefont
  {Saha-Dasgupta}, \citenamefont {Held},\ and\ \citenamefont
  {Capone}}]{Valli2016}%
  \BibitemOpen
  \bibfield  {author} {\bibinfo {author} {\bibfnamefont {A.}~\bibnamefont
  {Valli}}, \bibinfo {author} {\bibfnamefont {A.}~\bibnamefont {Amaricci}},
  \bibinfo {author} {\bibfnamefont {A.}~\bibnamefont {Toschi}}, \bibinfo
  {author} {\bibfnamefont {T.}~\bibnamefont {Saha-Dasgupta}}, \bibinfo {author}
  {\bibfnamefont {K.}~\bibnamefont {Held}}, \ and\ \bibinfo {author}
  {\bibfnamefont {M.}~\bibnamefont {Capone}},\ }\bibfield  {title} {\enquote
  {\bibinfo {title} {Effective magnetic correlations in hole-doped graphene
  nanoflakes},}\ }\href {\doibase 10.1103/PhysRevB.94.245146} {\bibfield
  {journal} {\bibinfo  {journal} {Phys. Rev. B}\ }\textbf {\bibinfo {volume}
  {94}},\ \bibinfo {pages} {245146} (\bibinfo {year} {2016})}\BibitemShut
  {NoStop}%
\bibitem [{\citenamefont {Valli}\ \emph {et~al.}(2018)\citenamefont {Valli},
  \citenamefont {Amaricci}, \citenamefont {Brosco},\ and\ \citenamefont
  {Capone}}]{Valli2018}%
  \BibitemOpen
  \bibfield  {author} {\bibinfo {author} {\bibfnamefont {A.}~\bibnamefont
  {Valli}}, \bibinfo {author} {\bibfnamefont {A.}~\bibnamefont {Amaricci}},
  \bibinfo {author} {\bibfnamefont {V.}~\bibnamefont {Brosco}}, \ and\ \bibinfo
  {author} {\bibfnamefont {M.}~\bibnamefont {Capone}},\ }\bibfield  {title}
  {\enquote {\bibinfo {title} {Quantum interference assisted spin filtering in
  graphene nanoflakes},}\ }\href {\doibase 10.1021/acs.nanolett.8b00453}
  {\bibfield  {journal} {\bibinfo  {journal} {Nano Lett.}\ }\textbf {\bibinfo
  {volume} {18}},\ \bibinfo {pages} {2158} (\bibinfo {year}
  {2018})}\BibitemShut {NoStop}%
\bibitem [{\citenamefont {Valli}\ \emph {et~al.}(2019)\citenamefont {Valli},
  \citenamefont {Amaricci}, \citenamefont {Brosco},\ and\ \citenamefont
  {Capone}}]{Valli2019}%
  \BibitemOpen
  \bibfield  {author} {\bibinfo {author} {\bibfnamefont {A.}~\bibnamefont
  {Valli}}, \bibinfo {author} {\bibfnamefont {A.}~\bibnamefont {Amaricci}},
  \bibinfo {author} {\bibfnamefont {V.}~\bibnamefont {Brosco}}, \ and\ \bibinfo
  {author} {\bibfnamefont {M.}~\bibnamefont {Capone}},\ }\bibfield  {title}
  {\enquote {\bibinfo {title} {Interplay between destructive quantum
  interference and symmetry-breaking phenomena in graphene quantum
  junctions},}\ }\href {\doibase 10.1103/PhysRevB.100.075118} {\bibfield
  {journal} {\bibinfo  {journal} {Phys. Rev. B}\ }\textbf {\bibinfo {volume}
  {100}},\ \bibinfo {pages} {075118} (\bibinfo {year} {2019})}\BibitemShut
  {NoStop}%
\bibitem [{\citenamefont {Meisner}\ \emph {et~al.}(2012)\citenamefont
  {Meisner}, \citenamefont {Ahn}, \citenamefont {Aradhya}, \citenamefont
  {Krikorian}, \citenamefont {Parameswaran}, \citenamefont {Steigerwald},
  \citenamefont {Venkataraman},\ and\ \citenamefont {Nuckolls}}]{Meisner2012}%
  \BibitemOpen
  \bibfield  {author} {\bibinfo {author} {\bibfnamefont {J.~S.}\ \bibnamefont
  {Meisner}}, \bibinfo {author} {\bibfnamefont {S.}~\bibnamefont {Ahn}},
  \bibinfo {author} {\bibfnamefont {S.~V.}\ \bibnamefont {Aradhya}}, \bibinfo
  {author} {\bibfnamefont {M.}~\bibnamefont {Krikorian}}, \bibinfo {author}
  {\bibfnamefont {R.}~\bibnamefont {Parameswaran}}, \bibinfo {author}
  {\bibfnamefont {M.}~\bibnamefont {Steigerwald}}, \bibinfo {author}
  {\bibfnamefont {L.}~\bibnamefont {Venkataraman}}, \ and\ \bibinfo {author}
  {\bibfnamefont {C.}~\bibnamefont {Nuckolls}},\ }\bibfield  {title} {\enquote
  {\bibinfo {title} {Importance of direct metal-$\pi$ coupling in electronic
  transport through conjugated single-molecule junctions},}\ }\href {\doibase
  doi: 10.1021/ja308626m} {\bibfield  {journal} {\bibinfo  {journal} {J. Am.
  Chem. Soc.}\ }\textbf {\bibinfo {volume} {134}},\ \bibinfo {pages} {20440}
  (\bibinfo {year} {2012})}\BibitemShut {NoStop}%
\bibitem [{\citenamefont {Manrique}\ \emph {et~al.}(2015)\citenamefont
  {Manrique}, \citenamefont {Huang}, \citenamefont {Baghernejad}, \citenamefont
  {Zhao}, \citenamefont {Al-Owaedi}, \citenamefont {Sadeghi}, \citenamefont
  {Kaliginedi}, \citenamefont {Hong}, \citenamefont {Gulcur}, \citenamefont
  {Wandlowski}, \citenamefont {Bryce},\ and\ \citenamefont
  {Lambert}}]{Manrique2015}%
  \BibitemOpen
  \bibfield  {author} {\bibinfo {author} {\bibfnamefont {D.~Z.}\ \bibnamefont
  {Manrique}}, \bibinfo {author} {\bibfnamefont {C.}~\bibnamefont {Huang}},
  \bibinfo {author} {\bibfnamefont {M.}~\bibnamefont {Baghernejad}}, \bibinfo
  {author} {\bibfnamefont {X.}~\bibnamefont {Zhao}}, \bibinfo {author}
  {\bibfnamefont {O.~A.}\ \bibnamefont {Al-Owaedi}}, \bibinfo {author}
  {\bibfnamefont {H.}~\bibnamefont {Sadeghi}}, \bibinfo {author} {\bibfnamefont
  {V.}~\bibnamefont {Kaliginedi}}, \bibinfo {author} {\bibfnamefont
  {W.}~\bibnamefont {Hong}}, \bibinfo {author} {\bibfnamefont {M.}~\bibnamefont
  {Gulcur}}, \bibinfo {author} {\bibfnamefont {T.}~\bibnamefont {Wandlowski}},
  \bibinfo {author} {\bibfnamefont {M.~R.}\ \bibnamefont {Bryce}}, \ and\
  \bibinfo {author} {\bibfnamefont {C.~J.}\ \bibnamefont {Lambert}},\
  }\bibfield  {title} {\enquote {\bibinfo {title} {A quantum circuit rule for
  interference effects in single-molecule electrical junctions},}\ }\href
  {\doibase 10.1038/ncomms7389} {\bibfield  {journal} {\bibinfo  {journal}
  {Nat. Commun.}\ }\textbf {\bibinfo {volume} {6}},\ \bibinfo {pages} {6389}
  (\bibinfo {year} {2015})}\BibitemShut {NoStop}%
\bibitem [{\citenamefont {Liu}\ \emph {et~al.}(2019)\citenamefont {Liu},
  \citenamefont {Huang}, \citenamefont {Wang},\ and\ \citenamefont
  {Hong}}]{Wang2019}%
  \BibitemOpen
  \bibfield  {author} {\bibinfo {author} {\bibfnamefont {J.}~\bibnamefont
  {Liu}}, \bibinfo {author} {\bibfnamefont {X.}~\bibnamefont {Huang}}, \bibinfo
  {author} {\bibfnamefont {F.}~\bibnamefont {Wang}}, \ and\ \bibinfo {author}
  {\bibfnamefont {W.}~\bibnamefont {Hong}},\ }\bibfield  {title} {\enquote
  {\bibinfo {title} {Quantum interference effects in charge transport through
  single-molecule junctions: Detection, manipulation, and application},}\
  }\href {\doibase 10.1021/acs.accounts.8b00429} {\bibfield  {journal}
  {\bibinfo  {journal} {Acc. Chem. Res.}\ }\textbf {\bibinfo {volume} {52}},\
  \bibinfo {pages} {151} (\bibinfo {year} {2019})}\BibitemShut {NoStop}%
\bibitem [{\citenamefont {Su}\ \emph {et~al.}(1979)\citenamefont {Su},
  \citenamefont {Schrieffer},\ and\ \citenamefont {Heeger}}]{Schrieffer1979}%
  \BibitemOpen
  \bibfield  {author} {\bibinfo {author} {\bibfnamefont {W.~P.}\ \bibnamefont
  {Su}}, \bibinfo {author} {\bibfnamefont {J.~R.}\ \bibnamefont {Schrieffer}},
  \ and\ \bibinfo {author} {\bibfnamefont {A.~J.}\ \bibnamefont {Heeger}},\
  }\bibfield  {title} {\enquote {\bibinfo {title} {Solitons in
  polyacetylene},}\ }\href {\doibase 10.1103/PhysRevLett.42.1698} {\bibfield
  {journal} {\bibinfo  {journal} {Phys. Rev. Lett.}\ }\textbf {\bibinfo
  {volume} {42}},\ \bibinfo {pages} {1698} (\bibinfo {year}
  {1979})}\BibitemShut {NoStop}%
\bibitem [{\citenamefont {Kuroda}\ and\ \citenamefont
  {Shirakawa}(1987)}]{Kuroda1987}%
  \BibitemOpen
  \bibfield  {author} {\bibinfo {author} {\bibfnamefont {S.-I.}\ \bibnamefont
  {Kuroda}}\ and\ \bibinfo {author} {\bibfnamefont {H.}~\bibnamefont
  {Shirakawa}},\ }\bibfield  {title} {\enquote {\bibinfo {title}
  {Electron-nuclear double-resonance evidence for the soliton wave function in
  polyacetylene},}\ }\href {\doibase 10.1103/PhysRevB.35.9380} {\bibfield
  {journal} {\bibinfo  {journal} {Phys. Rev. B}\ }\textbf {\bibinfo {volume}
  {35}},\ \bibinfo {pages} {9380(R)} (\bibinfo {year} {1987})}\BibitemShut
  {NoStop}%
\bibitem [{\citenamefont {Bally}\ \emph {et~al.}(2000)\citenamefont {Bally},
  \citenamefont {Hrovat},\ and\ \citenamefont {Borden}}]{Bally2000}%
  \BibitemOpen
  \bibfield  {author} {\bibinfo {author} {\bibfnamefont {T.}~\bibnamefont
  {Bally}}, \bibinfo {author} {\bibfnamefont {D.~A.}\ \bibnamefont {Hrovat}}, \
  and\ \bibinfo {author} {\bibfnamefont {W.~T.}\ \bibnamefont {Borden}},\
  }\bibfield  {title} {\enquote {\bibinfo {title} {Attempts to model neutral
  solitons in polyacetylene by {\em ab initio} and density functional methods.
  {T}he nature of the spin distribution in polyenyl radicals},}\ }\href
  {\doibase 10.1039/B003288N} {\bibfield  {journal} {\bibinfo  {journal} {Phys.
  Chem. Chem. Phys.}\ }\textbf {\bibinfo {volume} {2}},\ \bibinfo {pages}
  {3363} (\bibinfo {year} {2000})}\BibitemShut {NoStop}%
\bibitem [{\citenamefont {Wehling}\ \emph {et~al.}(2011)\citenamefont
  {Wehling}, \citenamefont {\c{S}a\c{s}io\v{g}lu}, \citenamefont {Friedrich},
  \citenamefont {Lichtenstein}, \citenamefont {Katsnelson},\ and\ \citenamefont
  {Bl\"ugel}}]{Wehling2011}%
  \BibitemOpen
  \bibfield  {author} {\bibinfo {author} {\bibfnamefont {T.~O.}\ \bibnamefont
  {Wehling}}, \bibinfo {author} {\bibfnamefont {E.}~\bibnamefont
  {\c{S}a\c{s}io\v{g}lu}}, \bibinfo {author} {\bibfnamefont {C.}~\bibnamefont
  {Friedrich}}, \bibinfo {author} {\bibfnamefont {A.~I.}\ \bibnamefont
  {Lichtenstein}}, \bibinfo {author} {\bibfnamefont {M.~I.}\ \bibnamefont
  {Katsnelson}}, \ and\ \bibinfo {author} {\bibfnamefont {S.}~\bibnamefont
  {Bl\"ugel}},\ }\bibfield  {title} {\enquote {\bibinfo {title} {Strength of
  effective {C}oulomb interactions in graphene and graphite},}\ }\href
  {\doibase 10.1103/PhysRevLett.106.236805} {\bibfield  {journal} {\bibinfo
  {journal} {Phys. Rev. Lett.}\ }\textbf {\bibinfo {volume} {106}},\ \bibinfo
  {pages} {236805} (\bibinfo {year} {2011})}\BibitemShut {NoStop}%
\bibitem [{\citenamefont {Datta}(2005)}]{Datta2005}%
  \BibitemOpen
  \bibfield  {author} {\bibinfo {author} {\bibfnamefont {S.}~\bibnamefont
  {Datta}},\ }\href {\doibase 10.1017/CBO9781139164313} {\emph {\bibinfo
  {title} {Quantum Transport: Atom to Transistor}}}\ (\bibinfo  {publisher}
  {Cambridge University Press},\ \bibinfo {year} {2005})\BibitemShut {NoStop}%
\bibitem [{\citenamefont {Feldner}\ \emph {et~al.}(2011)\citenamefont
  {Feldner}, \citenamefont {Meng}, \citenamefont {Lang}, \citenamefont
  {Assaad}, \citenamefont {Wessel},\ and\ \citenamefont
  {Honecker}}]{Feldner2011}%
  \BibitemOpen
  \bibfield  {author} {\bibinfo {author} {\bibfnamefont {H.}~\bibnamefont
  {Feldner}}, \bibinfo {author} {\bibfnamefont {Z.~Y.}\ \bibnamefont {Meng}},
  \bibinfo {author} {\bibfnamefont {T.~C.}\ \bibnamefont {Lang}}, \bibinfo
  {author} {\bibfnamefont {F.~F.}\ \bibnamefont {Assaad}}, \bibinfo {author}
  {\bibfnamefont {S.}~\bibnamefont {Wessel}}, \ and\ \bibinfo {author}
  {\bibfnamefont {A.}~\bibnamefont {Honecker}},\ }\bibfield  {title} {\enquote
  {\bibinfo {title} {Dynamical signatures of edge-state magnetism on graphene
  nanoribbons},}\ }\href {\doibase 10.1103/PhysRevLett.106.226401} {\bibfield
  {journal} {\bibinfo  {journal} {Phys. Rev. Lett.}\ }\textbf {\bibinfo
  {volume} {106}},\ \bibinfo {pages} {226401} (\bibinfo {year}
  {2011})}\BibitemShut {NoStop}%
\bibitem [{\citenamefont {Sorella}\ and\ \citenamefont
  {Tosatti}(1992)}]{Sorella1992}%
  \BibitemOpen
  \bibfield  {author} {\bibinfo {author} {\bibfnamefont {S.}~\bibnamefont
  {Sorella}}\ and\ \bibinfo {author} {\bibfnamefont {E.}~\bibnamefont
  {Tosatti}},\ }\bibfield  {title} {\enquote {\bibinfo {title}
  {Semi-metal-insulator transition of the {H}ubbard model in the honeycomb
  lattice},}\ }\href {\doibase 10.1209/0295-5075/19/8/007} {\bibfield
  {journal} {\bibinfo  {journal} {Europhys. Lett.}\ }\textbf {\bibinfo {volume}
  {19}},\ \bibinfo {pages} {699} (\bibinfo {year} {1992})}\BibitemShut
  {NoStop}%
\bibitem [{\citenamefont {Sorella}\ \emph {et~al.}(2012)\citenamefont
  {Sorella}, \citenamefont {Otsuka},\ and\ \citenamefont
  {Yunoki}}]{Sorella2012}%
  \BibitemOpen
  \bibfield  {author} {\bibinfo {author} {\bibfnamefont {S.}~\bibnamefont
  {Sorella}}, \bibinfo {author} {\bibfnamefont {Y.}~\bibnamefont {Otsuka}}, \
  and\ \bibinfo {author} {\bibfnamefont {S.}~\bibnamefont {Yunoki}},\
  }\bibfield  {title} {\enquote {\bibinfo {title} {Absence of a spin liquid
  phase in the {H}ubbard model on the honeycomb lattice},}\ }\href {\doibase
  10.1038/srep00992} {\bibfield  {journal} {\bibinfo  {journal} {Sci. Rep.}\
  }\textbf {\bibinfo {volume} {2}},\ \bibinfo {pages} {992} (\bibinfo {year}
  {2012})}\BibitemShut {NoStop}%
\bibitem [{\citenamefont {Hassan}\ and\ \citenamefont
  {S\'en\'echal}(2013)}]{Hassan2013}%
  \BibitemOpen
  \bibfield  {author} {\bibinfo {author} {\bibfnamefont {S.~R.}\ \bibnamefont
  {Hassan}}\ and\ \bibinfo {author} {\bibfnamefont {D.}~\bibnamefont
  {S\'en\'echal}},\ }\bibfield  {title} {\enquote {\bibinfo {title} {Absence of
  spin liquid in nonfrustrated correlated systems},}\ }\href {\doibase
  10.1103/PhysRevLett.110.096402} {\bibfield  {journal} {\bibinfo  {journal}
  {Phys. Rev. Lett.}\ }\textbf {\bibinfo {volume} {110}},\ \bibinfo {pages}
  {096402} (\bibinfo {year} {2013})}\BibitemShut {NoStop}%
\bibitem [{\citenamefont {Assaad}\ and\ \citenamefont
  {Herbut}(2013)}]{Assaad2013}%
  \BibitemOpen
  \bibfield  {author} {\bibinfo {author} {\bibfnamefont {F.~F.}\ \bibnamefont
  {Assaad}}\ and\ \bibinfo {author} {\bibfnamefont {I.~F.}\ \bibnamefont
  {Herbut}},\ }\bibfield  {title} {\enquote {\bibinfo {title} {Pinning the
  order: The nature of quantum criticality in the {H}ubbard model on honeycomb
  lattice},}\ }\href {\doibase 10.1103/PhysRevX.3.031010} {\bibfield  {journal}
  {\bibinfo  {journal} {Phys. Rev. X}\ }\textbf {\bibinfo {volume} {3}},\
  \bibinfo {pages} {031010} (\bibinfo {year} {2013})}\BibitemShut {NoStop}%
\bibitem [{\citenamefont {Hirschmeier}\ \emph {et~al.}(2018)\citenamefont
  {Hirschmeier}, \citenamefont {Hafermann},\ and\ \citenamefont
  {Lichtenstein}}]{Hirschmeier2018}%
  \BibitemOpen
  \bibfield  {author} {\bibinfo {author} {\bibfnamefont {D.}~\bibnamefont
  {Hirschmeier}}, \bibinfo {author} {\bibfnamefont {H.}~\bibnamefont
  {Hafermann}}, \ and\ \bibinfo {author} {\bibfnamefont {A.~I.}\ \bibnamefont
  {Lichtenstein}},\ }\bibfield  {title} {\enquote {\bibinfo {title} {Multiband
  dual fermion approach to quantum criticality in the {H}ubbard honeycomb
  lattice},}\ }\href {\doibase 10.1103/PhysRevB.97.115150} {\bibfield
  {journal} {\bibinfo  {journal} {Phys. Rev. B}\ }\textbf {\bibinfo {volume}
  {97}},\ \bibinfo {pages} {115150} (\bibinfo {year} {2018})}\BibitemShut
  {NoStop}%
\bibitem [{\citenamefont {Raczkowski}\ \emph {et~al.}(2020)\citenamefont
  {Raczkowski}, \citenamefont {Peters}, \citenamefont {Ph\`ung}, \citenamefont
  {Takemori}, \citenamefont {Assaad}, \citenamefont {Honecker},\ and\
  \citenamefont {Vahedi}}]{Marcin2019}%
  \BibitemOpen
  \bibfield  {author} {\bibinfo {author} {\bibfnamefont {M.}~\bibnamefont
  {Raczkowski}}, \bibinfo {author} {\bibfnamefont {R.}~\bibnamefont {Peters}},
  \bibinfo {author} {\bibfnamefont {T.~T.}\ \bibnamefont {Ph\`ung}}, \bibinfo
  {author} {\bibfnamefont {N.}~\bibnamefont {Takemori}}, \bibinfo {author}
  {\bibfnamefont {F.~F.}\ \bibnamefont {Assaad}}, \bibinfo {author}
  {\bibfnamefont {A.}~\bibnamefont {Honecker}}, \ and\ \bibinfo {author}
  {\bibfnamefont {J.}~\bibnamefont {Vahedi}},\ }\bibfield  {title} {\enquote
  {\bibinfo {title} {Hubbard model on the honeycomb lattice: From static and
  dynamical mean-field theories to lattice quantum {M}onte {C}arlo
  simulations},}\ }\href {\doibase 10.1103/PhysRevB.101.125103} {\bibfield
  {journal} {\bibinfo  {journal} {Phys. Rev. B}\ }\textbf {\bibinfo {volume}
  {101}},\ \bibinfo {pages} {125103} (\bibinfo {year} {2020})}\BibitemShut
  {NoStop}%
\bibitem [{\citenamefont {Georges}\ \emph {et~al.}(1996)\citenamefont
  {Georges}, \citenamefont {Kotliar}, \citenamefont {Krauth},\ and\
  \citenamefont {Rozenberg}}]{Georges1996}%
  \BibitemOpen
  \bibfield  {author} {\bibinfo {author} {\bibfnamefont {A.}~\bibnamefont
  {Georges}}, \bibinfo {author} {\bibfnamefont {G.}~\bibnamefont {Kotliar}},
  \bibinfo {author} {\bibfnamefont {W.}~\bibnamefont {Krauth}}, \ and\ \bibinfo
  {author} {\bibfnamefont {M.~J.}\ \bibnamefont {Rozenberg}},\ }\bibfield
  {title} {\enquote {\bibinfo {title} {Dynamical mean-field theory of strongly
  correlated fermion systems and the limit of infinite dimensions},}\ }\href
  {\doibase 10.1103/RevModPhys.68.13} {\bibfield  {journal} {\bibinfo
  {journal} {Rev. Mod. Phys.}\ }\textbf {\bibinfo {volume} {68}},\ \bibinfo
  {pages} {13} (\bibinfo {year} {1996})}\BibitemShut {NoStop}%
\bibitem [{\citenamefont {Wilson}(1975)}]{Wilson1975}%
  \BibitemOpen
  \bibfield  {author} {\bibinfo {author} {\bibfnamefont {K.~G.}\ \bibnamefont
  {Wilson}},\ }\bibfield  {title} {\enquote {\bibinfo {title} {The
  renormalization group: {C}ritical phenomena and the {K}ondo problem},}\
  }\href {\doibase 10.1103/RevModPhys.47.773} {\bibfield  {journal} {\bibinfo
  {journal} {Rev. Mod. Phys.}\ }\textbf {\bibinfo {volume} {47}},\ \bibinfo
  {pages} {773} (\bibinfo {year} {1975})}\BibitemShut {NoStop}%
\bibitem [{\citenamefont {Krishna-murthy}\ \emph {et~al.}(1980)\citenamefont
  {Krishna-murthy}, \citenamefont {Wilkins},\ and\ \citenamefont
  {Wilson}}]{Krishna1980}%
  \BibitemOpen
  \bibfield  {author} {\bibinfo {author} {\bibfnamefont {H.~R.}\ \bibnamefont
  {Krishna-murthy}}, \bibinfo {author} {\bibfnamefont {J.~W.}\ \bibnamefont
  {Wilkins}}, \ and\ \bibinfo {author} {\bibfnamefont {K.~G.}\ \bibnamefont
  {Wilson}},\ }\bibfield  {title} {\enquote {\bibinfo {title}
  {Renormalization-group approach to the {A}nderson model of dilute magnetic
  alloys. {I}. {S}tatic properties for the symmetric case},}\ }\href {\doibase
  10.1103/PhysRevB.21.1003} {\bibfield  {journal} {\bibinfo  {journal} {Phys.
  Rev. B}\ }\textbf {\bibinfo {volume} {21}},\ \bibinfo {pages} {1003}
  (\bibinfo {year} {1980})}\BibitemShut {NoStop}%
\bibitem [{\citenamefont {Bulla}\ \emph {et~al.}(2008)\citenamefont {Bulla},
  \citenamefont {Costi},\ and\ \citenamefont {Pruschke}}]{Bulla2008}%
  \BibitemOpen
  \bibfield  {author} {\bibinfo {author} {\bibfnamefont {R.}~\bibnamefont
  {Bulla}}, \bibinfo {author} {\bibfnamefont {T.~A.}\ \bibnamefont {Costi}}, \
  and\ \bibinfo {author} {\bibfnamefont {T.}~\bibnamefont {Pruschke}},\
  }\bibfield  {title} {\enquote {\bibinfo {title} {Numerical renormalization
  group method for quantum impurity systems},}\ }\href {\doibase
  10.1103/RevModPhys.80.395} {\bibfield  {journal} {\bibinfo  {journal} {Rev.
  Mod. Phys.}\ }\textbf {\bibinfo {volume} {80}},\ \bibinfo {pages} {395}
  (\bibinfo {year} {2008})}\BibitemShut {NoStop}%
\bibitem [{\citenamefont {Peters}\ and\ \citenamefont
  {Kawakami}(2014)}]{Robert2014}%
  \BibitemOpen
  \bibfield  {author} {\bibinfo {author} {\bibfnamefont {R.}~\bibnamefont
  {Peters}}\ and\ \bibinfo {author} {\bibfnamefont {N.}~\bibnamefont
  {Kawakami}},\ }\bibfield  {title} {\enquote {\bibinfo {title} {Spin density
  waves in the {H}ubbard model: A {DMFT} approach},}\ }\href {\doibase
  10.1103/PhysRevB.89.155134} {\bibfield  {journal} {\bibinfo  {journal} {Phys.
  Rev. B}\ }\textbf {\bibinfo {volume} {89}},\ \bibinfo {pages} {155134}
  (\bibinfo {year} {2014})}\BibitemShut {NoStop}%
\bibitem [{\citenamefont {Peters}\ and\ \citenamefont
  {Kawakami}(2015)}]{Robert2015}%
  \BibitemOpen
  \bibfield  {author} {\bibinfo {author} {\bibfnamefont {R.}~\bibnamefont
  {Peters}}\ and\ \bibinfo {author} {\bibfnamefont {N.}~\bibnamefont
  {Kawakami}},\ }\bibfield  {title} {\enquote {\bibinfo {title} {Large and
  small {F}ermi-surface spin density waves in the {K}ondo lattice model},}\
  }\href {\doibase 10.1103/PhysRevB.92.075103} {\bibfield  {journal} {\bibinfo
  {journal} {Phys. Rev. B}\ }\textbf {\bibinfo {volume} {92}},\ \bibinfo
  {pages} {075103} (\bibinfo {year} {2015})}\BibitemShut {NoStop}%
\bibitem [{\citenamefont {Peters}\ \emph {et~al.}(2006)\citenamefont {Peters},
  \citenamefont {Pruschke},\ and\ \citenamefont {Anders}}]{Robert2006}%
  \BibitemOpen
  \bibfield  {author} {\bibinfo {author} {\bibfnamefont {R.}~\bibnamefont
  {Peters}}, \bibinfo {author} {\bibfnamefont {T.}~\bibnamefont {Pruschke}}, \
  and\ \bibinfo {author} {\bibfnamefont {F.~B.}\ \bibnamefont {Anders}},\
  }\bibfield  {title} {\enquote {\bibinfo {title} {Numerical renormalization
  group approach to {G}reen's functions for quantum impurity models},}\ }\href
  {\doibase 10.1103/PhysRevB.74.245114} {\bibfield  {journal} {\bibinfo
  {journal} {Phys. Rev. B}\ }\textbf {\bibinfo {volume} {74}},\ \bibinfo
  {pages} {245114} (\bibinfo {year} {2006})}\BibitemShut {NoStop}%
\bibitem [{\citenamefont {Ph\`ung}(2019)}]{phdThu}%
  \BibitemOpen
  \bibfield  {author} {\bibinfo {author} {\bibfnamefont {T.~T.}\ \bibnamefont
  {Ph\`ung}},\ }\emph {\bibinfo {title} {Numerical Studies of Magnetism and
  Transport Properties in Graphene Nano-Devices}},\ \href
  {https://hal.archives-ouvertes.fr/tel-02631069} {Ph.D. thesis},\ \bibinfo
  {school} {Universit\'e de Cergy-Pontoise} (\bibinfo {year}
  {2019})\BibitemShut {NoStop}%
\bibitem [{\citenamefont {Sierra}\ \emph {et~al.}(2018)\citenamefont {Sierra},
  \citenamefont {Neumann}, \citenamefont {Cuppens}, \citenamefont {Raes},
  \citenamefont {Costache},\ and\ \citenamefont {Valenzuela}}]{Sierra2018}%
  \BibitemOpen
  \bibfield  {author} {\bibinfo {author} {\bibfnamefont {J.~F.}\ \bibnamefont
  {Sierra}}, \bibinfo {author} {\bibfnamefont {I.}~\bibnamefont {Neumann}},
  \bibinfo {author} {\bibfnamefont {J.}~\bibnamefont {Cuppens}}, \bibinfo
  {author} {\bibfnamefont {B.}~\bibnamefont {Raes}}, \bibinfo {author}
  {\bibfnamefont {M.~V.}\ \bibnamefont {Costache}}, \ and\ \bibinfo {author}
  {\bibfnamefont {S.~O.}\ \bibnamefont {Valenzuela}},\ }\bibfield  {title}
  {\enquote {\bibinfo {title} {Thermoelectric spin voltage in graphene},}\
  }\href {\doibase 10.1038/s41565-017-0015-9} {\bibfield  {journal} {\bibinfo
  {journal} {Nat. Nanotechnol.}\ }\textbf {\bibinfo {volume} {13}},\ \bibinfo
  {pages} {107} (\bibinfo {year} {2018})}\BibitemShut {NoStop}%
\bibitem [{\citenamefont {Tang}\ \emph {et~al.}(2018)\citenamefont {Tang},
  \citenamefont {Ye}, \citenamefont {Tan},\ and\ \citenamefont
  {Ren}}]{Tang2018}%
  \BibitemOpen
  \bibfield  {author} {\bibinfo {author} {\bibfnamefont {X.-Q.}\ \bibnamefont
  {Tang}}, \bibinfo {author} {\bibfnamefont {X.-M.}\ \bibnamefont {Ye}},
  \bibinfo {author} {\bibfnamefont {X.-Y.}\ \bibnamefont {Tan}}, \ and\
  \bibinfo {author} {\bibfnamefont {D.-H.}\ \bibnamefont {Ren}},\ }\bibfield
  {title} {\enquote {\bibinfo {title} {Metal-free magnetism, spin-dependent
  {S}eebeck effect, and spin-{S}eebeck diode effect in armchair graphene
  nanoribbons},}\ }\href {\doibase 10.1038/s41598-018-19632-3} {\bibfield
  {journal} {\bibinfo  {journal} {Sci. Rep.}\ }\textbf {\bibinfo {volume}
  {8}},\ \bibinfo {pages} {927} (\bibinfo {year} {2018})}\BibitemShut {NoStop}%
\bibitem [{\citenamefont {Borges}\ \emph {et~al.}(2017)\citenamefont {Borges},
  \citenamefont {Xia}, \citenamefont {Liu}, \citenamefont {Venkataraman},\ and\
  \citenamefont {Solomon}}]{Borges2017}%
  \BibitemOpen
  \bibfield  {author} {\bibinfo {author} {\bibfnamefont {A.}~\bibnamefont
  {Borges}}, \bibinfo {author} {\bibfnamefont {J.}~\bibnamefont {Xia}},
  \bibinfo {author} {\bibfnamefont {S.~H.}\ \bibnamefont {Liu}}, \bibinfo
  {author} {\bibfnamefont {L.}~\bibnamefont {Venkataraman}}, \ and\ \bibinfo
  {author} {\bibfnamefont {G.~C.}\ \bibnamefont {Solomon}},\ }\bibfield
  {title} {\enquote {\bibinfo {title} {The role of through-space interactions
  in modulating constructive and destructive interference effects in
  benzene},}\ }\href {\doibase 10.1021/acs.nanolett.7b01592} {\bibfield
  {journal} {\bibinfo  {journal} {Nano Lett.}\ }\textbf {\bibinfo {volume}
  {17}},\ \bibinfo {pages} {4436} (\bibinfo {year} {2017})}\BibitemShut
  {NoStop}%
\bibitem [{\citenamefont {Hikihara}\ \emph {et~al.}(2003)\citenamefont
  {Hikihara}, \citenamefont {Hu}, \citenamefont {Lin},\ and\ \citenamefont
  {Mou}}]{Hikihara2003}%
  \BibitemOpen
  \bibfield  {author} {\bibinfo {author} {\bibfnamefont {T.}~\bibnamefont
  {Hikihara}}, \bibinfo {author} {\bibfnamefont {X.}~\bibnamefont {Hu}},
  \bibinfo {author} {\bibfnamefont {H.-H.}\ \bibnamefont {Lin}}, \ and\
  \bibinfo {author} {\bibfnamefont {C.-Y.}\ \bibnamefont {Mou}},\ }\bibfield
  {title} {\enquote {\bibinfo {title} {Ground-state properties of nanographite
  systems with zigzag edges},}\ }\href {\doibase 10.1103/PhysRevB.68.035432}
  {\bibfield  {journal} {\bibinfo  {journal} {Phys. Rev. B}\ }\textbf {\bibinfo
  {volume} {68}},\ \bibinfo {pages} {035432} (\bibinfo {year}
  {2003})}\BibitemShut {NoStop}%
\bibitem [{\citenamefont {Son}\ \emph {et~al.}(2006)\citenamefont {Son},
  \citenamefont {Cohen},\ and\ \citenamefont {Louie}}]{Son2006}%
  \BibitemOpen
  \bibfield  {author} {\bibinfo {author} {\bibfnamefont {Y.-W.}\ \bibnamefont
  {Son}}, \bibinfo {author} {\bibfnamefont {M.~L.}\ \bibnamefont {Cohen}}, \
  and\ \bibinfo {author} {\bibfnamefont {S.~G.}\ \bibnamefont {Louie}},\
  }\bibfield  {title} {\enquote {\bibinfo {title} {Half-metallic graphene
  nanoribbons},}\ }\href {\doibase 10.1038/nature05180} {\bibfield  {journal}
  {\bibinfo  {journal} {Nature}\ }\textbf {\bibinfo {volume} {444}},\ \bibinfo
  {pages} {347} (\bibinfo {year} {2006})}\BibitemShut {NoStop}%
\bibitem [{\citenamefont {Fern\'andez-Rossier}(2008)}]{Fernandez2008}%
  \BibitemOpen
  \bibfield  {author} {\bibinfo {author} {\bibfnamefont {J.}~\bibnamefont
  {Fern\'andez-Rossier}},\ }\bibfield  {title} {\enquote {\bibinfo {title}
  {Prediction of hidden multiferroic order in graphene zigzag ribbons},}\
  }\href {\doibase 10.1103/PhysRevB.77.075430} {\bibfield  {journal} {\bibinfo
  {journal} {Phys. Rev. B}\ }\textbf {\bibinfo {volume} {77}},\ \bibinfo
  {pages} {075430} (\bibinfo {year} {2008})}\BibitemShut {NoStop}%
\bibitem [{\citenamefont {Magda}\ \emph {et~al.}(2014)\citenamefont {Magda},
  \citenamefont {Jin}, \citenamefont {Hagym\'asi}, \citenamefont {Vancs\'o},
  \citenamefont {Osv\'ath}, \citenamefont {Nemes-Incze}, \citenamefont {Hwang},
  \citenamefont {Bir\'o},\ and\ \citenamefont {Tapaszt\'o}}]{Magda2014}%
  \BibitemOpen
  \bibfield  {author} {\bibinfo {author} {\bibfnamefont {G.~Z.}\ \bibnamefont
  {Magda}}, \bibinfo {author} {\bibfnamefont {X.}~\bibnamefont {Jin}}, \bibinfo
  {author} {\bibfnamefont {I.}~\bibnamefont {Hagym\'asi}}, \bibinfo {author}
  {\bibfnamefont {P.}~\bibnamefont {Vancs\'o}}, \bibinfo {author}
  {\bibfnamefont {Z.}~\bibnamefont {Osv\'ath}}, \bibinfo {author}
  {\bibfnamefont {P.}~\bibnamefont {Nemes-Incze}}, \bibinfo {author}
  {\bibfnamefont {C.}~\bibnamefont {Hwang}}, \bibinfo {author} {\bibfnamefont
  {L.~P.}\ \bibnamefont {Bir\'o}}, \ and\ \bibinfo {author} {\bibfnamefont
  {L.}~\bibnamefont {Tapaszt\'o}},\ }\bibfield  {title} {\enquote {\bibinfo
  {title} {Room-temperature magnetic order on zigzag edges of narrow graphene
  nanoribbons},}\ }\href {\doibase 10.1038/nature13831} {\bibfield  {journal}
  {\bibinfo  {journal} {Nature}\ }\textbf {\bibinfo {volume} {514}},\ \bibinfo
  {pages} {608} (\bibinfo {year} {2014})}\BibitemShut {NoStop}%
\bibitem [{\citenamefont {Chen}\ \emph {et~al.}(2017)\citenamefont {Chen},
  \citenamefont {Zhou}, \citenamefont {Yu}, \citenamefont {Yin},\ and\
  \citenamefont {Gong}}]{Chen2017}%
  \BibitemOpen
  \bibfield  {author} {\bibinfo {author} {\bibfnamefont {W.-C.}\ \bibnamefont
  {Chen}}, \bibinfo {author} {\bibfnamefont {Y.}~\bibnamefont {Zhou}}, \bibinfo
  {author} {\bibfnamefont {S.-L.}\ \bibnamefont {Yu}}, \bibinfo {author}
  {\bibfnamefont {W.-G.}\ \bibnamefont {Yin}}, \ and\ \bibinfo {author}
  {\bibfnamefont {C.-D.}\ \bibnamefont {Gong}},\ }\bibfield  {title} {\enquote
  {\bibinfo {title} {Width-tuned magnetic order oscillation on zigzag edges of
  honeycomb nanoribbons},}\ }\href {\doibase 10.1021/acs.nanolett.7b01474}
  {\bibfield  {journal} {\bibinfo  {journal} {Nano Lett.}\ }\textbf {\bibinfo
  {volume} {17}},\ \bibinfo {pages} {4400} (\bibinfo {year}
  {2017})}\BibitemShut {NoStop}%
\bibitem [{\citenamefont {Zhai}\ \emph {et~al.}(2019)\citenamefont {Zhai},
  \citenamefont {Gu}, \citenamefont {Wen}, \citenamefont {Liu}, \citenamefont
  {Zhu}, \citenamefont {Zhou}, \citenamefont {Gong},\ and\ \citenamefont
  {Li}}]{Zhai2019}%
  \BibitemOpen
  \bibfield  {author} {\bibinfo {author} {\bibfnamefont {X.}~\bibnamefont
  {Zhai}}, \bibinfo {author} {\bibfnamefont {J.}~\bibnamefont {Gu}}, \bibinfo
  {author} {\bibfnamefont {R.}~\bibnamefont {Wen}}, \bibinfo {author}
  {\bibfnamefont {R.-W.}\ \bibnamefont {Liu}}, \bibinfo {author} {\bibfnamefont
  {M.}~\bibnamefont {Zhu}}, \bibinfo {author} {\bibfnamefont {X.}~\bibnamefont
  {Zhou}}, \bibinfo {author} {\bibfnamefont {L.-Y.}\ \bibnamefont {Gong}}, \
  and\ \bibinfo {author} {\bibfnamefont {X.}~\bibnamefont {Li}},\ }\bibfield
  {title} {\enquote {\bibinfo {title} {Giant {S}eebeck magnetoresistance
  triggered by electric field and assisted by a valley through a
  ferromagnetic/antiferromagnetic junction in heavy group-{IV} monolayers},}\
  }\href {\doibase 10.1103/PhysRevB.99.085421} {\bibfield  {journal} {\bibinfo
  {journal} {Phys. Rev. B}\ }\textbf {\bibinfo {volume} {99}},\ \bibinfo
  {pages} {085421} (\bibinfo {year} {2019})}\BibitemShut {NoStop}%
\bibitem [{\citenamefont {Lv}\ and\ \citenamefont {Zhao}(2019)}]{LV2019}%
  \BibitemOpen
  \bibfield  {author} {\bibinfo {author} {\bibfnamefont {Y.-Z.}\ \bibnamefont
  {Lv}}\ and\ \bibinfo {author} {\bibfnamefont {P.}~\bibnamefont {Zhao}},\
  }\bibfield  {title} {\enquote {\bibinfo {title} {Spin caloritronic transport
  of tree-saw graphene nanoribbons},}\ }\href {\doibase
  10.1088/0256-307X/36/1/017301} {\bibfield  {journal} {\bibinfo  {journal}
  {Chin. Phys. Lett.}\ }\textbf {\bibinfo {volume} {36}},\ \bibinfo {pages}
  {017301} (\bibinfo {year} {2019})}\BibitemShut {NoStop}%
\bibitem [{\citenamefont {Tan}\ \emph {et~al.}(2018)\citenamefont {Tan},
  \citenamefont {Wu}, \citenamefont {Liu}, \citenamefont {Fu},\ and\
  \citenamefont {Wu}}]{Tan2018}%
  \BibitemOpen
  \bibfield  {author} {\bibinfo {author} {\bibfnamefont {X.-Y.}\ \bibnamefont
  {Tan}}, \bibinfo {author} {\bibfnamefont {D.-D.}\ \bibnamefont {Wu}},
  \bibinfo {author} {\bibfnamefont {Q.-B.}\ \bibnamefont {Liu}}, \bibinfo
  {author} {\bibfnamefont {H.-H.}\ \bibnamefont {Fu}}, \ and\ \bibinfo {author}
  {\bibfnamefont {R.}~\bibnamefont {Wu}},\ }\bibfield  {title} {\enquote
  {\bibinfo {title} {Spin caloritronics in armchair silicene nanoribbons with
  $sp^{3}$ and $sp^{2}$-type alternating hybridizations},}\ }\href {\doibase
  10.1088/1361-648x/aad4b9} {\bibfield  {journal} {\bibinfo  {journal} {J.
  Phys.: Condens. Matter}\ }\textbf {\bibinfo {volume} {30}},\ \bibinfo {pages}
  {355303} (\bibinfo {year} {2018})}\BibitemShut {NoStop}%
\bibitem [{\citenamefont {Guimar\~aes}\ \emph {et~al.}(2014)\citenamefont
  {Guimar\~aes}, \citenamefont {van~den Berg}, \citenamefont {Vera-Marun},
  \citenamefont {Zomer},\ and\ \citenamefont {van Wees}}]{Guimaraes2014}%
  \BibitemOpen
  \bibfield  {author} {\bibinfo {author} {\bibfnamefont {M.~H.~D.}\
  \bibnamefont {Guimar\~aes}}, \bibinfo {author} {\bibfnamefont {J.~J.}\
  \bibnamefont {van~den Berg}}, \bibinfo {author} {\bibfnamefont {I.~J.}\
  \bibnamefont {Vera-Marun}}, \bibinfo {author} {\bibfnamefont {P.~J.}\
  \bibnamefont {Zomer}}, \ and\ \bibinfo {author} {\bibfnamefont {B.~J.}\
  \bibnamefont {van Wees}},\ }\bibfield  {title} {\enquote {\bibinfo {title}
  {Spin transport in graphene nanostructures},}\ }\href {\doibase
  10.1103/PhysRevB.90.235428} {\bibfield  {journal} {\bibinfo  {journal} {Phys.
  Rev. B}\ }\textbf {\bibinfo {volume} {90}},\ \bibinfo {pages} {235428}
  (\bibinfo {year} {2014})}\BibitemShut {NoStop}%
\end{thebibliography}%

\end{document}